# The Rise of Zirconium Metal–Organic Polyhedra

Joohan Nam[†,1,2], Seohyeon Kim[†,2], Eunjie Ju[2], Shinhyeok Moon[3] and Wonyoung Choe[*,2,3,4,5]

[1]*Chair of Inorganic Chemistry I, Technische Universität Dresden (TU Dresden), Bergstraße 66, 01069 Dresden, Germany.*

[2]*Department of Chemistry,* [3]*Graduate School of Carbon Neutrality,* [4]*Department of Mechanical Engineering, and* [5]*Graduate School of Artificial Intelligence, Ulsan National Institute of Science and Technology (UNIST), 50 UNIST-gil, 44919 Ulsan, Republic of Korea.*

[†]These authors contributed equally to this work.

*Correspondence: choe@unist.ac.kr

## Abstract

Zirconium metal–organic polyhedra (Zr-MOPs) assembled from cyclopentadienyl (Cp)-capped $Zr_3$ clusters and carboxylate linkers form a structurally diverse family. Which MOP forms depends on the connectivity of the $Zr_3$ node, the geometry of the linker, and assembly conditions. This review classifies Zr-MOPs into edge- and face-directed tetrahedra, a cube, and cocoons according to the connectivity and geometry of their building blocks. From reported structures, we map which linkers give which MOPs and examine how the $Zr_3$ cluster precursor and the reaction conditions decide between the tetrahedron and the cocoon when one linker can give both. We describe how hydrogen bonding and ionic association organize the MOPs into reticular nets and porous salts in the crystal. Together, these comparisons relate cage geometry to node–linker connectivity and crystal packing to the geometry and distribution of external interaction sites, providing a structural basis for designing Zr-MOPs and their supramolecular assemblies.

## 1. Introduction

Polyhedra were studied as geometric objects long before chemists learned to construct them as molecules.[1] The quest to synthesize polyhedral molecules has led to landmark achievements, including cubane,[2] tetrahedrane derivatives,[3] and dodecahedrane.[4] Related geometries found in nature include the $B_{12}$ icosahedra of boron-rich solids[5] and the 24-subunit protein shell of ferritin.[6] Metal–organic polyhedra (MOPs) are discrete molecular cages, unlike metal–organic frameworks (MOFs), which form extended frameworks.[7–11] These cages provide a modular approach to constructing molecular polyhedra by assembling metal nodes and organic linkers with appropriate connectivity and geometry.[12] Our 2021 survey presented zirconium MOPs (Zr-MOPs), assembled from cyclopentadienyl (Cp)-capped $Zr_3$ clusters and carboxylate linkers, as an emerging cage family.[13] Subsequent studies have extended their use to catalysis, gas separation, desalination, and biomedicine.[14,15]

The structural chemistry of Zr-MOPs begins with a common $Zr_3$ node. The oxophilicity of Zr(IV) supports robust Zr–carboxylate connections,[14–16] and the $Cp_3Zr_3(\mu_3\text{-O})(\mu_2\text{-OH})_3$ cluster presents three carboxylate binding sites on the side of the $Zr_3$ plane opposite the Cp rings. Post-assembly reactions can modify the linkers while preserving the cage scaffold, allowing chemical modifications to be compared within a defined molecular structure.[17–19] Each cage also has an internal cavity and an external surface.[20,21] Intermolecular contacts involving this surface organize the cages in the crystal.[22–26] Molecular identity and crystal packing therefore require separate structural evidence.[27,28] The geometry that closes a cage and the interactions that assemble it in a crystal present two related design questions: how do the building blocks form a discrete molecule, and how do these molecules organize in the solid state?

In this review, we focus on discrete MOPs assembled from $Zr_3$ nodes and their noncovalent organization in the solid state. Our earlier review established a topology-based classification across different metal nodes.[13] For Zr-MOPs, complementary perspectives have related cage–anion hydrogen bonding to reticular nets[25] and examined connections between discrete cages and extended materials.[16] Accounts of synthesis, functionalization, and applications[14] have further explored the relationship between controlled assembly and function.[15] These studies provide a foundation for examining the structural diversity of Zr-MOPs. Interestingly, a common node, and sometimes the same linker, can give rise to different cages, while a given cage can adopt different packing arrangements. Understanding this diversity requires molecular closure and intermolecular association to be considered together.

This review has two goals. First, we classify the reported Zr-MOPs as edge- and face-directed tetrahedra, a cube, and cocoons according to building-block connectivity and geometry. We map the linker–cage relationships and examine how precursor chemistry and reaction conditions influence tetrahedron–cocoon selection and interconversion. Second, we examine how hydroxo-group orientation, molecular anions, and other intermolecular contacts organize these cages in the solid state. Through these comparisons, we aim to identify the variables that govern molecular identity and crystal packing and provide a basis for directing assembly at both structural levels.

## 2. Classification of Polyhedra

We describe the polyhedral geometries of Zr-MOPs using the three-letter codes of the Reticular Chemistry Structure Resource (RCSR) database (Figure 1).[29] Each polyhedron is characterized by its numbers of vertices, edges, and faces and by its transitivity ($p$,$q$,$r$), the numbers of symmetry-distinct vertices, edges, and faces (Table 1). Both polyhedra covered here, the tetrahedron (**tet**) and the cube (**cub**), are regular, with transitivity (1,1,1).

In these polyhedral representations, $Zr_3$ nodes or branch points of linkers serve as vertices, and the connections between them define the edges. In edge-directed $V_4E_6$ tetrahedra, linkers span the six edges, whereas in face-directed $V_4F_4$ tetrahedra, they span the four faces (V, vertex; E, edge; F, face; Figure 1b, d). For the face-directed tetrahedron, treating the branch points of the tritopic linkers as additional vertices gives a metal–linker bipartite graph corresponding to the heterocube (**cub-b**) (Figure 1d).[13,30]

Beyond these polyhedral structures, the same $Zr_3$ nodes also form cocoons, in which two nodes are joined by three ditopic linkers; these cocoons are introduced in Section 4.3. This cage has been described as a lantern,[31,32] a cigar[14,27] or a capsule;[15,33] we call it a cocoon throughout this review.

## 3. $Zr_3$ Nodes and Linkers for Zirconium MOP Assembly

### 3.1. $Zr_3$ Clusters and Structural Notation

The Cp-capped $Zr_3$ cluster provides three carboxylate binding sites and is treated as a single three-connected $Zr_3$ node (Figure 2a).[30] Four $Zr_3$ clusters contain twelve Zr atoms but contribute four cage vertices to the polyhedral representation. Cp substitution changes the peripheral chemistry of this unit (Figure 2b). Besides the parent Cp, *n*-butyl-, benzyl-, 4-(trifluoromethyl)benzyl- and 4-vinylbenzyl-substituted Cp rings have been used,[34,35] and pentamethylcyclopentadienyl (Cp*) caps appear in cocoons.[36]

Linkers are grouped according to their geometry, connectivity, and backbone structure: A for linear and zigzag ditopic linkers, B for bent ditopic linkers, C for tritopic linkers, and D for metal-bridged linkers (Figures 3–12).

### 3.2. Linear Ditopic Linkers

Linkers based on benzene-1,4-dicarboxylic acid ($H_2BDC$) form the largest family (Figure 3).[30,37] Substituted derivatives carry one or more substituents, including amino, hydroxy, alkyl, halo, azido, unsaturated, alkoxy, and heterocyclic groups, while benzo-fused and pyridine analogs modify the ring itself.[36–67] BDC derivatives also include amide-, imine-, and triazole-functionalized linkers and metal-binding derivatives bearing an isatin-derived imine–carbonyl chelating site (**H₂A28**) or its Co, Mn, and Ni complexes (**H₂A31**–**H₂A33**).[39,68–76]

[1,1′-Biphenyl]-4,4′-dicarboxylic acid ($H_2$BPDC) and [1,1′:4′,1″-terphenyl]-4,4″-dicarboxylic acid ($H_2$TPDC) extend the distance between the carboxylate groups and provide additional ring positions for functionalization (Figures 4 and 5).[27,37,77–79] A related nitrogen-containing backbone is [2,2′-bipyridine]-5,5′-dicarboxylic acid ($H_2$bpydc, **$H_2$A58**), with metal-coordinated derivatives (**$H_2$A59**–**$H_2$A64**, Figure 4).[80–85]

For $H_2$TPDC, the central ring can also be replaced by naphthalene, anthracene, or benzimidazolium.[27,41] Functionalization after MOP assembly has also been reported, including DNA attachment through azide–dibenzocyclooctyne (DBCO) coupling (**$H_2$A69**, **$H_2$A71**).[36]

[1,1′:4′,1″:4″,1‴-Quaterphenyl]-4,4‴-dicarboxylic acid ($H_2$QPDC) and related linkers carry two benzoate arms on a central biaryl core (Figure 6).[77] Among these derivatives, linkers with a central binaphthyl unit exhibit axial chirality,[86,87] and those bearing diphenylphosphanyl substituents provide coordination sites for Ni(II) (**$H_2$A84**) or Cu(I) (**$H_2$A85**).[86] Nitrogen-containing cores include a Ru-coordinated 2,2′-bipyridine unit (**$H_2$A87**) and a 4,4′-bipyridinium unit (**$H_2$A86**).[83,88]

Other linear backbones include Schiff-base linkers incorporating Cr, Mn, or Fe complexes (Figure 7a) and a boron-dipyrromethene (BODIPY) linker containing a chromophoric core

(Figure 7b).[89,90] Anthracene-based backbones are also represented by two linkers in this group.[27,91]

### 3.3. Zigzag Ditopic Linkers

Zigzag-shaped linkers keep their two carboxylate axes parallel but laterally offset (Figure 8). This arrangement can arise from aromatic substitution patterns (**H₂A94**), *trans* C=C, N=N, or C=N linkages, or nonaromatic backbone segments (**$H_2$A95**, **$H_2$A96**).[27,77,92,93] Further variations include bipyridinium, thiazolothiazole, imine, azine, and sulfonated ether-linked units.[33,94–96] This group also includes cyclic backbones containing phosphorus and nitrogen (**$H_2$A109**–**$H_2$A113**).[97] Several of these linkers (**$H_2$A96**, **$H_2$A103**, **$H_2$A106**–**$H_2$A113**) contain $sp^3$ chains, methylene hinges, or an eight-membered ring between the carboxylates, which allow rotation about single bonds. These flexible linkers are drawn in extended conformations.

### 3.4. Bent Ditopic Linkers

The bent ditopic linkers considered here vary in backbone geometry, flexibility, and the positions of their coordinating groups (Figure 9). The 1,3-substitution pattern distinguishes the isophthalic acid derivatives (**$H_2$B1**, **$H_2$B6**) from the linear 1,4-substituted BDC family.[98] Representative backbones include norbornene-derived rings (**$H_2$B2**, **$H_2$B3**), methylene-linked pyrazole units (**$H_2$B4**, **$H_2$B5**), methylene-, sulfone-, and ether-linked aromatic units (**$H_2$B7**–**$H_2$B9**), and diphenylethene-, thiophene-, and indole-based structures (**$H_2$B10**, **$H_2$B11**, **$H_2$B13**, **$H_2$B20**).[27,32,92,99–102] Squaramide-containing backbones are also included (**$H_2$B21**).[103] A central imidazolium ring connects two carboxylate-bearing aryl arms through its 1,3-positions, forming a bent backbone with a cationic core (**$H_2$B14–$H_2$B17**).[104,105]

### 3.5. Tritopic Linkers

The tritopic linker family includes benzene-1,3,5-tricarboxylic acid ($H_3$BTC, **$H_3$C1**), its extended analog bearing three benzoate arms (**$H_3$C3**), and their substituted or further extended derivatives (Figure 10). Other backbones incorporate central nitrogen or boron atoms, amide linkages, or charged units. Tritopic connectivity does not require three-fold symmetry, and asymmetric linker backbones have also been reported (Figure 11).[30,94,106–114]

### 3.6. Metal-Bridged Linkers

In some linkers, carboxylate groups coordinate $Zr_3$ nodes, while nitrogen-donor sites bind an additional metal center, allowing separate organic components to assemble into a metal-bridged linker.[115] A Pd(II) center, for example, bridges two pyridyl termini to form a ditopic linker (Figure 12). Metal bridging also gives tritopic linkers. Three Cu centers bind three pyrazolate groups to form a triangular assembly bearing three outward-facing carboxylate groups (**$H_3$D6**).[114] In this review, each metal-bridged assembly is treated as a single linker, and its internal metal centers are not counted as separate metal nodes.

## 4. Cage Structures

The reported Zr-MOPs comprise two regular polyhedral types, the tetrahedron (**tet**) and the cube (**cub**), as well as nonpolyhedral cocoons. The compositions and linkers reported for each cage type are summarized in Table 2.

### 4.1. Tetrahedron (tet)

The tetrahedron is the simplest regular polyhedron, with four vertices, six edges and four triangular faces. Zr-based tetrahedra adopt either edge-directed or face-directed structures. In edge-directed tetrahedra, six ditopic linkers connect four $Zr_3$ nodes along the tetrahedral edges. In face-directed tetrahedra, four tritopic linkers each connect the three $Zr_3$ nodes defining a triangular face. The corresponding compositions are $M_4L_6$ and $M_4L_4$, respectively, where M denotes a $Zr_3$ cluster node and L a linker.[30]

**Edge-directed $M_4L_6$ tetrahedra.** BDC-, BPDC-, and TPDC-based tetrahedra form an isoreticular series in which longer aromatic backbones increase the separation between $Zr_3$ nodes while preserving the tetrahedral connectivity (Figures 13–15).[30,37,116,117] Ring substitution modifies the local environment around these edges. In the methyl-substituted BDC cage (**$H_2$A2**), the substituents narrow the windows while leaving the internal cavity essentially unchanged in size relative to the unsubstituted cage (Figure 13j).[118] In the bipyridine-based tetrahedron (**$H_2$A58**), the linker-based *N*,*N*-chelating sites remain uncoordinated (Figure 14e). These sites can coordinate additional metal centers along the tetrahedral edges while preserving the cage connectivity, as illustrated by the $PdCl_2$- and $CuCl_2$-coordinated derivatives (Figure 14a, b).[80]

Salen-based linkers form tetrahedra in which Mn(III) or Cr(III) occupies the midpoint of each edge (**$H_2$A90**, **$H_2$A88**; Figure 16b, c). The two metal-containing cages are isostructural. The enantiopure cyclohexanediamine backbones retain their stereochemistry within the cages, resulting in homochiral tetrahedral structures.[89]

Edge-directed tetrahedra also accommodate zigzag and bent linker backbones (Figures 17 and 18). Fumarate-, naphthalene-2,6-dicarboxylate-, and azobenzene-based linkers form $M_4L_6$ tetrahedra (**$H_2$A95**, **$H_2$A94**, **$H_2$A101**; Figure 17a–c).[106,119,120] In the azobenzene-based tetrahedron, the linker adopts a nearly straight conformation (Figure 17c).[120] Despite its bent backbone, the Au(I)–*N*-heterocyclic carbene linker connects $Zr_3$ nodes into an $M_4L_6$ tetrahedron (**$H_2$B19**; Figure 18).[104]

**Face-directed $M_4L_4$ tetrahedra.** The BTC-based tetrahedron contains four tritopic linkers, each connecting the three $Zr_3$ nodes that define a triangular face (Figure 19a).[30] Extension of the linker arms through phenyl and ethynyl units, or with biphenyl units around a nitrogen center, enlarges the faces while preserving the $M_4L_4$ connectivity (Figure 19d–f).[30,106,110] Metal-bridged assemblies can also occupy these face positions. In one example, three Cu centers joined by pyrazolate groups form a triangular linker whose three outward-facing carboxylates connect to the $Zr_3$ nodes (Figure 19c).[114]

Tetrahedral connectivity is retained with tritopic linkers that lack three-fold symmetry, but the cages depart from ideal tetrahedral geometry (Figure 20). Isohedral tetrahedra have symmetry-equivalent faces but two sets of edge lengths (Figure 20a, b).[109] A carbazole-based linker gives a tetrahedral cage with $D_{2d}$ symmetry (Figure 20c).[110]

## 4.2. Cube (cub)

A cube has eight vertices, twelve edges and six square faces. The reported $M_8L_{12}$ cage contains eight $Zr_3$ nodes connected by twelve camphorate linkers along the cubic edges (Figure 21). The chiral linkers give a homochiral cage, and the crystal is also chiral.[121]

## 4.3. Cocoons

$M_2L_3$ cocoons are nonpolyhedral cages in which two $Zr_3$ nodes face each other and are connected by three ditopic linkers.[27,31,77] Each node connects through all three linkers to the same partner, rather than to three different nodes as in an edge-directed tetrahedron. The cocoon has the same 2:3 node-to-linker ratio but half the molecular mass of an $M_4L_6$ tetrahedron built from the same components.[35]

The $M_2L_3$ connectivity is retained across cocoons with different linker lengths and backbone geometries (Figures 22–27). Within the BDC-, BPDC-, and TPDC-based series, isoreticular expansion increases the separation between the two $Zr_3$ nodes.[27,122,123] The bound linkers can also undergo substantial bending. Short, rigid BDC backbones bend out of plane in compact cocoons, while the azobenzene-based zigzag linker adopts a bent conformation despite being nearly straight in its tetrahedral counterpart (Figures 22 and 26f).[27,120] Changes within the phosphorus-containing zigzag linker also alter the cage outline. Upon $PdCl_2$ or $PtCl_2$ coordination, the metal-binding units face outward and the central rings change conformation, making the cocoon longer and slimmer than the parent structure (Figure 27a, c, d).[97] In the methylene-linked viologen example, the three linkers wind around the axis connecting the two nodes and form a triple helix (Figure 27e).[94]

In several cocoons built from bent linkers, the central portions of the linkers lie around the cage equator, with their arms extending toward the two terminal $Zr_3$ nodes (Figures 28 and 29). The imidazolium-based examples have an overall molecular shape resembling a rugby ball (Figure 29h, j).[104]

In Pd-bridged cocoons, carboxylate–pyridyl linkers bind $Zr_3$ nodes through their carboxylate groups and Pd(II) through their pyridyl nitrogen atoms (Figures 30 and 31).[115,124] In the *cis*-Pd series, Pd(II) units join the pyridyl termini of two preformed $Zr_3$ clusters to close the cage, placing three Pd centers around its equatorial region (Figure 31).[115] Longer pyridyl–benzoate arms enlarge the cage while retaining its $M_2L_3$ connectivity (Figure 31c). The ancillary ligands at Pd also affect the windows, which are larger in the ethylenediamine-containing cage than in its *N*-methylated analog (Figure 31b, a).[115]

### 4.4. Cp-Substituted Cages

Cp-ring substituents project from the face of the $Zr_3$ cluster opposite the carboxylate sites, decorating the cage exterior without forming part of the metal–linker connections (Figures 32–34).[125,126] Crystallographic analysis confirmed that the benzyl-Cp tetrahedron retains cluster geometry and edge dimensions comparable to those of the parent cage (Figure 32e).[35] Benzyl-type substitution on the Cp rings decreases water solubility, while 4-vinylbenzyl and 4-(trifluoromethyl)benzyl substitution enhances solubility in *N*,*N*-dimethylformamide (DMF) (Figure 32b, d, e).[35] *n*-Butyl substitution on the Cp rings improves methanol solubility in a diamino-BPDC tetrahedron, enabling incorporation into cross-linked polymer membranes (Figure 32h).[127] Cp* caps were introduced to improve stability during DNA conjugation, and the azido-TPDC cocoon remained intact under the reported conjugation conditions (Figure 34e).[36]

## 5. Assembly Selection and Structural Interconversion

The same $Zr_3$ node and ditopic linker can close into either an $M_4L_6$ tetrahedron or an $M_2L_3$ cocoon.[27,31,77,128] Such pairs allow cage selection to be examined while retaining the principal building components. Table 2 and Figure 35 summarize the linkers reported for each cage type. Within related linker series, width relative to length provides an empirical measure of cage preference.[27] The close arrangement of three linkers in a cocoon limits the lateral bulk it can accommodate. **$H_2$BDC** (**$H_2$A1**) formed both cage types, whereas 2,5-dimethyl substitution (**$H_2$A34**) enabled isolation of a phase-pure tetrahedron. In the TPDC series, 2′,5′-dimethyl substitution (**$H_2$A74**) was insufficient to isolate a phase-pure tetrahedron, while isobutoxy substitution (**$H_2$A75**) enabled its isolation. Linker sterics thus provide a structural basis for cage preference within these series.

Preformed $Zr_3$ clusters introduce precursor chemistry as a variable in cage selection. Conventional synthesis from $Cp_2ZrCl_2$ combines $Zr_3$ cluster formation and cage assembly, often producing tetrahedron–cocoon mixtures. Preformed $Zr_3$ cluster precursors separate these steps: cage-forming linkers replace monocarboxylate capping ligands on a defined precursor.[129] Benzoate-capped $Zr_3$ cluster precursors afforded phase-pure cages from **$H_2$A1** and **$H_2$A52** under different conditions.[31] With **$H_2$A1**, water addition favored the cocoon from the chloride salt of the benzoate-capped $Zr_3$ cluster, whereas the triflate salt of the same cluster retained tetrahedral selectivity. Changing the capping ligand from benzoate to pivalate increased tetrahedral selectivity from 72% to 96% for **$H_2$A58** under the reported comparison conditions.[77] Reaction monitoring supported preferential tetrahedral formation, and bulky capping ligands were proposed to disfavor compact intermediates along the cocoon pathway. High selectivity was retained at the gram scale, although the outcome remained linker-dependent.[77] These results show that precursor chemistry can adjust selectivity while retaining the cage-forming linker.

Solubility extends precursor control to the conditions under which a cage can assemble. An acetate-capped $Zr_3$ cluster precursor, more soluble than its benzoate analog, enabled room-temperature assembly of phase-pure tetrahedra from ditopic and tritopic linkers.[106] Fumaric acid (**$H_2$A95**) gave a tetrahedron through this route, whereas conventional solvothermal synthesis had produced the cocoon. The different outcomes connect precursor choice and assembly conditions to cage geometry for an unchanged linker. Capping-ligand design and precursor solubility therefore provide complementary ways to extend assembly control beyond linker modification.

Reaction time provides another means of selecting among cages, with time-dependent product distributions and direct interconversion offering distinct evidence. With **$H_2$A52** and the triflate salt of the benzoate-capped $Zr_3$ cluster, a tetrahedron was isolated after 20 min, whereas cocoons were present at reaction times exceeding 30 min.[31] Time-resolved measurements directly followed conversion of the **$H_2$A101**-based tetrahedron to the cocoon. Temperature-dependent equilibrium studies in methanol supported an entropically driven process, with solvent release proposed to contribute.[120] The conversion response depends on the cage and solution conditions. Free pivalic acid promoted conversion of the isolated **$H_2$A58** tetrahedron to the cocoon in DMF,[77] whereas no detectable interconversion was observed for isolated **$H_2$A1** cages over 21 days in methanol at room temperature.[130] Purification can also alter the cage distribution in isolated fractions, separating formation selectivity from the purity of the recovered product.[130] Once selected and isolated, the molecular cage becomes a building block for assembly in the solid state.

## 6. Supramolecular Assembly of Polyhedral Units

In our previous work, we treated discrete Zr cages and hydrogen-bonded anionic clusters as complementary building blocks for supramolecular frameworks.[25] The geometry of their contacts connects molecular structure to packing topology. Edge-directed $M_4L_6$ and face-directed $M_4L_4$ tetrahedra both contain four $Zr_3$ nodes, but their $\mu_2$-OH groups point toward the tetrahedral edges and faces, respectively. When all three hydroxo groups at each node participate in chloride-mediated contacts, the edge-directed cage combines with an eight-connected anionic cluster to form a (4,8)-connected **flu** net. The face-directed cage instead combines with a four-connected anionic cluster to form a (4,4)-connected **dia-b** net (Figure 36a).[25] Both cages occupy four-connected sites. The distinct nets therefore reflect the distribution of interaction sites on cages with the same overall tetrahedral shape.

The anionic cluster makes the distinction between physical geometry and effective connectivity explicit. In **flu** and **dia-b**, cuboctahedral and octahedral chloride arrangements correspond to eight-connected cubic and four-connected tetrahedral patterns, respectively. Their physical geometries are related to a cube and a tetrahedron by rectification, which places vertices at the original edge midpoints.[25] Solvent participation changes the contacts available for assembly. In DMF-containing α frameworks, solvent binding leaves two hydroxo groups per $Zr_3$ node involved in chloride-mediated contacts. Two-connected anionic units then link either tetrahedral cage type into interpenetrated **dia** nets. In *N*,*N*-dimethylacetamide (DMA)-containing β frameworks, all three hydroxo groups participate, supporting **flu** or **dia-b** packing.[131] An $M_8L_{12}$ cubic cage also forms **flu**, with the cage at the eight-connected site and the anionic cluster at the four-connected site. The same net is thus obtained by exchanging the roles of the cage and anionic cluster.[25,121]

Known packing structures also provide a starting point for identifying candidate nets. Using this framework, we screened edge-transitive nets against building-block geometry and connectivity, then applied a reverse topological approach to identify the required cage type and hydroxo-group orientation.[25] The resulting models were proposed as geometrically compatible candidates for targeted synthesis, rather than experimentally realized structures in that study. Figure 36b summarizes these candidates together with the reported nets for each combination of cage and anionic cluster. This workflow connects net selection to the molecular configuration and interaction sites required for its construction.

The number of intermolecular connections can change even when the molecular cage is retained. **$H_2$A52**-based tetrahedra form **dia**, **bnn**, and **xhy** frameworks under different crystallization conditions.[116] In **dia**, chloride-mediated vertex-to-vertex contacts generate a 2-fold interpenetrated network. In **bnn**, obtained by changing the solvent mixture to *N*,*N*-diethylformamide, three $Zr_3$ nodes retain these connections, while the remaining vertex participates in vertex-to-face contacts involving chloride and aromatic C–H groups. Including these contacts gives a five-connected representation with trigonal-bipyramidal connection geometry. Chloride remains the bridging anion in both structures despite nitrate addition during synthesis. Recrystallization with excess nitrate instead affords **xhy**, where nitrate and methanol connect cages represented as eight-connected nodes.[116] These structures retain the same four-vertex cage, while the contacts used to define the supramolecular node determine its effective connectivity.

Changes in chloride-mediated contacts can also accompany variations in crystal symmetry. Triphenylamine-based **$H_3$C2** tetrahedra with $M_4L_4$ connectivity form centrosymmetric racemic, homochiral, and racemic noncentrosymmetric crystals under different crystallization conditions.[113] These packing differences were attributed to changes in cage–chloride hydrogen bonding upon HCl addition.[113]

The assembly partner can also be varied by pairing cationic Zr tetrahedra with molecular anions. Sulfonate-functionalized anionic Cu cages associate with Zr tetrahedra through hydroxo–sulfonate hydrogen bonds in the resolved crystals.[132] Rapid mixing produced an amorphous powder with a 6:1 Zr-to-Cu cage ratio, whereas slow layering afforded 2:1 cage crystals with additional charge-balancing cations. Further salt-metathesis studies varied reaction stoichiometry and addition rate to tune product composition and particle size, respectively.[133] Keggin-type polyoxometalates (POMs), including $[SiMo_{12}O_{40}]^{4-}$ and $[SiW_{12}O_{40}]^{4-}$, provide another class of molecular anions, contacting tetrahedral vertices or edges in crystals obtained by methanol diffusion into DMA.[134,135] Molecular-anion identity and preparation route therefore affect both composition and association. The resolved crystal structures describe specific products, as illustrated by the different powder X-ray diffraction patterns of POM-containing crystals and bulk precipitates.[134]

## 7. Packing of Cocoons

Our previous work analyzed packing topologies in polyhedral and polygonal Zr cages, leaving cocoon assemblies outside its scope.[25] For elongated $M_2L_3$ cocoons, intermolecular connectivity must be considered together with relative molecular orientation. Parallel arrangements are common,[122] and contacts at the terminal $Zr_3$ nodes and along the linkers connect the cages into extended structures. The **H$_2$A1**-, **H$_2$A100**-, and **H$_2$A101**-based cocoons form three-dimensional networks through hydrogen bonds involving chloride ions and solvent molecules.[136] Linker functionality adds further contacts. In the two-dimensional packing of the **H$_2$B8** cocoon, sulfonyl oxygen atoms contact the linker and Cp regions of neighboring cages through C–H···O=S and C···O=S interactions. The **$H_2$B7** analog retains the molecular geometry but adopts a different packing arrangement. The sulfonyl contacts were proposed to reinforce the packing of the **$H_2$B8** cocoon, linking the external chemistry of the cage to its intermolecular organization.[100]

The **$H_2$A54** cocoon shows how a common molecular unit can support different relative orientations. The nonparallel structure, obtained under 1,4-dioxane-based conditions, and the parallel structure, obtained under DMF-based conditions, retain the same $M_2L_3$ unit (Figure 23d).[122] In the nonparallel structure, oppositely oriented cocoons form pairs that assemble into axial motifs connected along three directions through chloride-mediated hydrogen bonds. The chloride ions define an octahedral assembly: each chloride accepts hydroxo hydrogen bonds from three cages, while two hydroxo groups from a cage are associated with an octahedral edge. The reported representation treats this assembly as a twelve-connected anionic cluster.[122] Its effective connectivity differs from the four-connected anionic cluster in **dia-b** despite their similar octahedral chloride geometries.[25] This nonparallel structure retained crystallinity after gas adsorption and exhibited accessible porosity.[122] These assemblies illustrate how cocoon packing couples relative cage orientation to the connectivity of the surrounding anionic assembly.

## 8. Conclusions and Future Perspectives

Zr-MOPs illustrate how a common $Zr_3$ node can give rise to different molecular cages and crystal assemblies. Different node–linker connections produce edge- and face-directed tetrahedra, a cube, and cocoons. Tetrahedron–cocoon pairs show how precursor capping ligands and reaction conditions influence cage selection for a given linker, while isolation procedures affect the purity of the recovered material. At the crystal level, hydroxo-group orientation, molecular anions, and other intermolecular contacts organize a given cage geometry into different packing arrangements. Cocoon assemblies further illustrate the role of relative molecular orientation. Across these examples, molecular connectivity defines cage closure, while the geometry and distribution of external interaction sites guide supramolecular assembly. Considering these two structural levels together provides a basis for directing both molecular structure and crystal packing.

The next step toward predictive assembly is to establish when molecular preferences become reproducible selection rules and which intermolecular contact patterns favor a particular packing net. Addressing these questions requires structure-resolved records that connect building-block identity with reaction conditions, isolation history, and characterization. Such records should specify precursor capping ligands and counteranions, solution composition and water content, temperature, reaction time, and isolation procedures. Molecular identity and crystal packing should be documented separately, with the supporting bulk or single-crystal evidence clearly specified. Unsuccessful syntheses should also be recorded where available. Computational screening and robotic synthesis of organic cages provide a precedent for linking candidate structures with experimental outcomes.[137] Combining these records with geometric screening would allow compatible candidate nets to be evaluated against experimentally established assembly outcomes.[25] The growing structural catalog could then become a source of testable predictions for designing molecular cages and their assembly in the solid state.

## Author Information

### Corresponding Author

*E-mail: choe@unist.ac.kr

### ORCID

Joohan Nam: 0000-0001-5123-418X

Seohyeon Kim: 0009-0002-3970-4382

Eunjie Ju: 0000-0002-7748-6681

Shinhyeok Moon: 0009-0007-4220-238X

Wonyoung Choe: 0000-0003-0957-1187

## Author contributions

All authors contributed to the writing, review, and revision of the manuscript.

## Conflict of Interest

The authors declare no competing financial interest.

## Data availability

No new data were generated in this review. The data discussed are available in the cited publications.

## Acknowledgements

This work was supported by National Research Foundation of Korea (NRF) grants funded by the Korean Government (NRF-2021M3I3A1084909, RS-2024-00463225, and RS-2025-00513741). This research was supported by Creative Materials Discovery Plus Program through the National Research Foundation of Korea (NRF) funded by Ministry of Science and ICT (RS-2026-25596523). This work was also supported by an Institute of Information & communications Technology Planning & Evaluation (IITP) grant funded by the Korea government (MSIT) (No. RS-2020-II201336, Artificial Intelligence Graduate School Program (UNIST)). J.N. acknowledges support from the Alexander von Humboldt Foundation through a Humboldt Research Fellowship.

**Table 1.** Basic structural information for the two polyhedra.

| Polyhedron | Type of vertex, *p* | Type of edge, *q* | Type of face, *r* | Number of vertices | Number of edges | Number of faces | Symmetry |
|---|---|---|---|---|---|---|---|
| **tet** | 1 | 1 | 1 | 4 | 6 | 4 | $T_d$ |
| **cub** | 1 | 1 | 1 | 8 | 12 | 6 | $O_h$ |

**Table 2.** Linkers used for each cage type of Zr-MOPs.

| Cage type | Composition | Linker shape | Linkers |
|---|---|---|---|
| **Tetrahedron** | Edge-directed $M_4L_6$ | 2-c linear | $H_2$A1–$H_2$A17, $H_3$A18, $H_2$A19–$H_2$A22, $H_2$A24–$H_2$A43, $H_4$A44, $H_2$A45–$H_2$A54, $H_2$A56–$H_2$A66, $H_2$A68, $H_2$A72–$H_2$A77, $H_2$A80–$H_2$A85, $H_2$A87–$H_2$A91 |
| | | 2-c zigzag | $H_2$A94, $H_2$A95, $H_2$A98–$H_2$A101 |
| | | 2-c bent (120°) | $H_2$B1, $H_2$B6 |
| | | 2-c bent (>120°) | $H_2$B18, $H_2$B19 |
| | Face-directed $M_4L_4$ | 3-c (120°) | $H_3$C1–$H_3$C11 |
| | | 3-c (distorted) | $H_3$C12–$H_3$C14 |
| | | 3-c (metal-bridged) | $H_3$D6 |
| **Cube** | $M_8L_{12}$ | 2-c bent (>120°) | $H_2$B12 |
| **Cocoon** | $M_2L_3$ | 2-c linear | $H_2$A1, $H_2$A12, $H_2$A14, $H_2$A23, $H_2$A25, $H_2$A50, $H_2$A52, $H_2$A54, $H_2$A55, $H_2$A58, $H_2$A66–$H_2$A71, $H_2$A74, $H_2$A76, $H_2$A78, $H_2$A79, $H_2$A86, $H_2$A92, $H_2$A93 |
| | | 2-c zigzag | $H_2$A94–$H_2$A98, $H_2$A100–$H_2$A113 |
| | | 2-c bent (<90°) | $H_2$B2, $H_2$B3 |
| | | 2-c bent (109°) | $H_2$B4, $H_2$B5, $H_2$B7–$H_2$B9 |
| | | 2-c bent (120°) | $H_2$B10, $H_2$B11 |
| | | 2-c bent (>120°) | $H_2$B13–$H_2$B17, $H_2$B19–$H_2$B21 |
| | | 2-c (metal-bridged) | $H_2$D1–$H_2$D5 |

The indicated angles denote the angles between the carboxylate groups expected from the linker backbone.

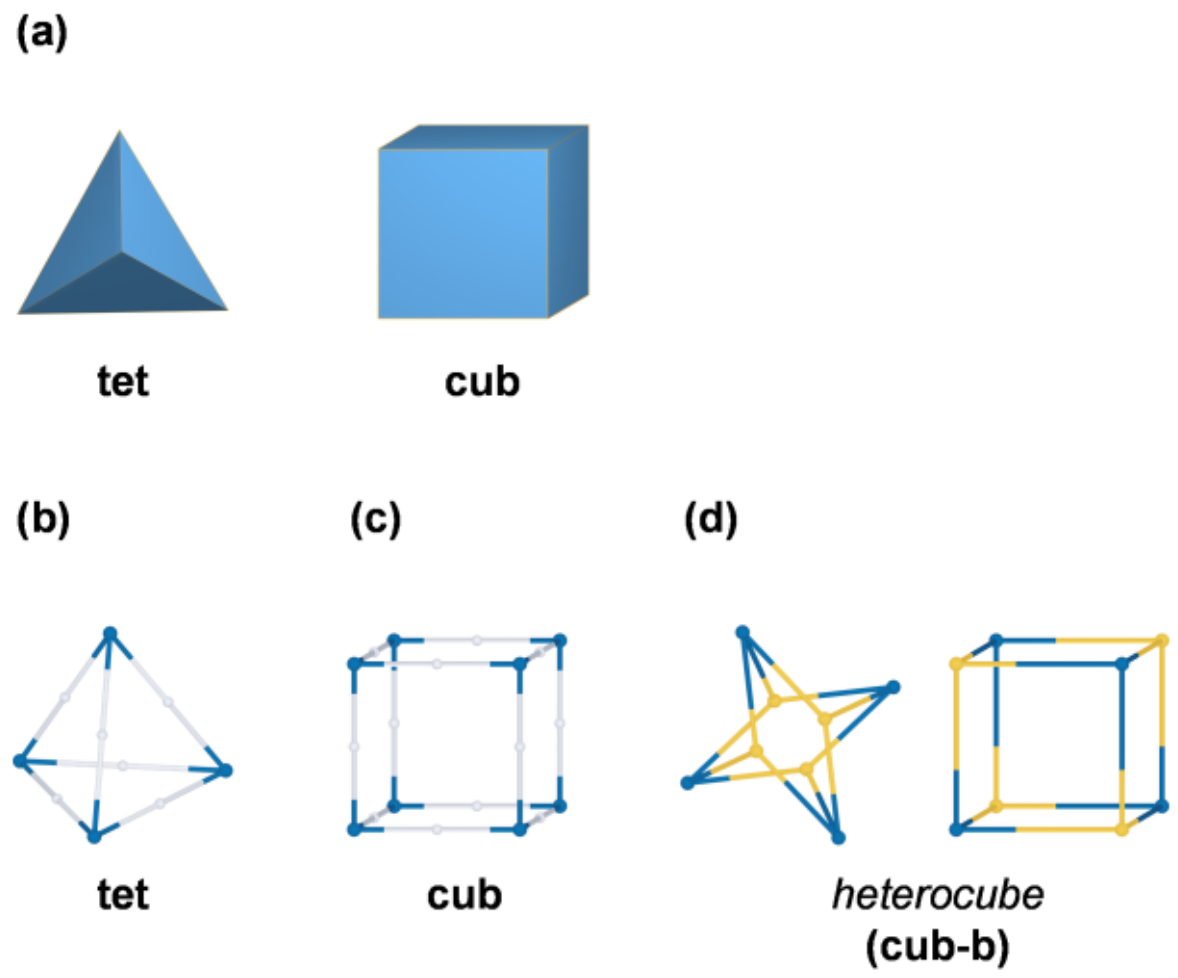

**Figure 1.** Description of the polyhedra covered in this review. The three boldfaced letters (**xyz**) are the RCSR codes for the polyhedra. (a) **tet** and **cub**. Skeletal views of (b) **tet**, (c) **cub**, and (d) the heterocube (**cub-b**). Adapted from "The rise of metal–organic polyhedra," S. Lee, H. Jeong, D. Nam, M. S. Lah and W. Choe, *Chem. Soc. Rev.*, 50, 528–555 (2021), with permission of the Royal Society of Chemistry; permission conveyed through Copyright Clearance Center, Inc.

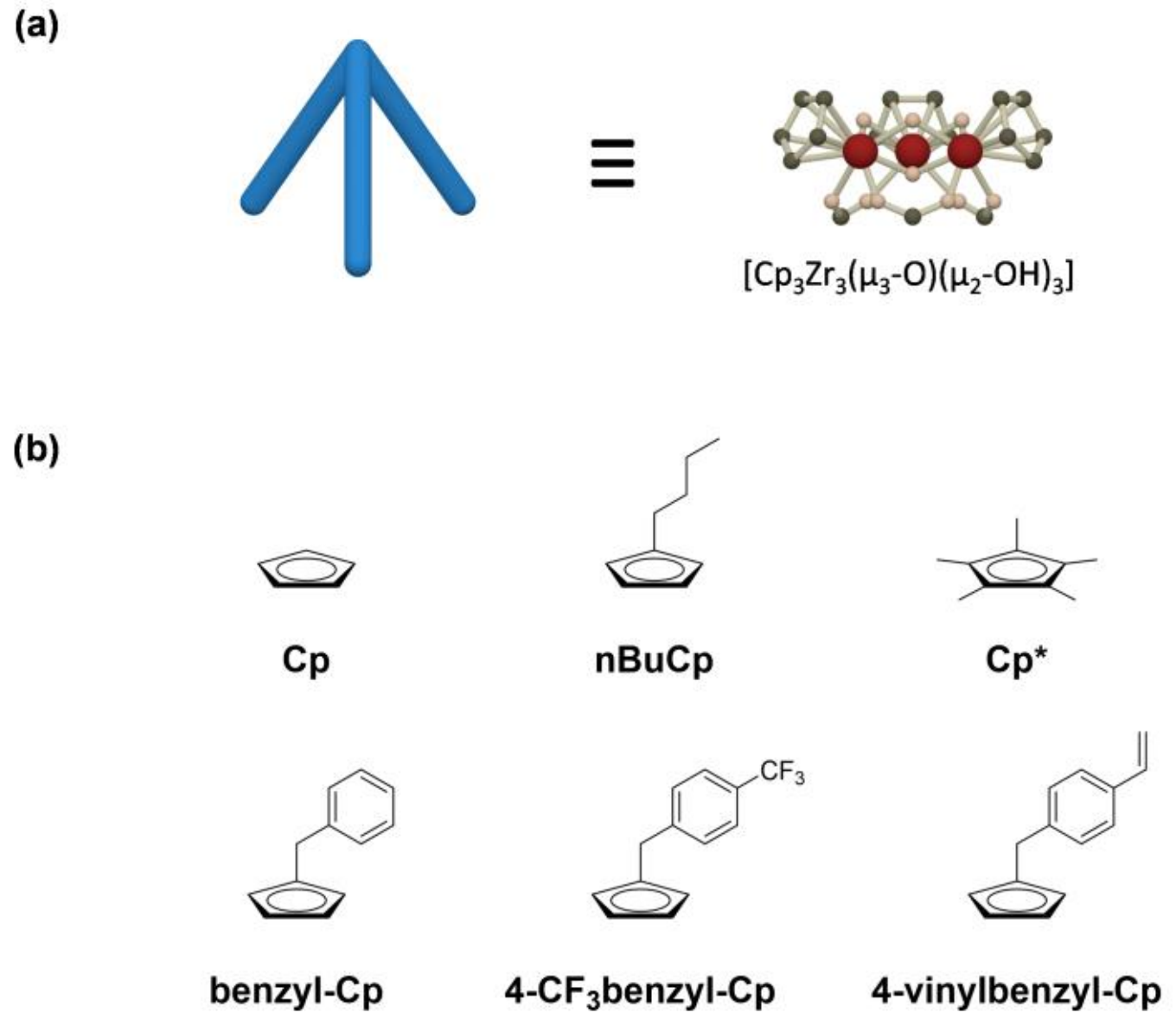

**Figure 2.** (a) The [$Cp_3Zr_3(\mu_3$-O)($\mu_2$-OH)$_3$] cluster with the three bridging carboxylate groups (O–C–O) of the linkers, and its representation as a three-connected $Zr_3$ node (blue). The three carboxylate sites lie on the side of the $Zr_3$ plane opposite to the Cp rings. (b) Cp caps found in Zr-MOPs: cyclopentadienyl (Cp), *n*-butylcyclopentadienyl (*n*BuCp), pentamethylcyclopentadienyl (Cp*), benzylcyclopentadienyl (benzyl-Cp), 4-(trifluoromethyl)benzylcyclopentadienyl (4-$CF_3$benzyl-Cp) and 4-vinylbenzylcyclopentadienyl (4-vinylbenzyl-Cp). Atom colors in (a): Zr, dark red; O, beige; C, dark gray. H atoms are omitted for clarity.[30,34–36] Panel (a) adapted from "The rise of metal–organic polyhedra," S. Lee, H. Jeong, D. Nam, M. S. Lah and W. Choe, *Chem. Soc. Rev.*, 50, 528–555 (2021), with permission of the Royal Society of Chemistry; permission conveyed through Copyright Clearance Center, Inc.

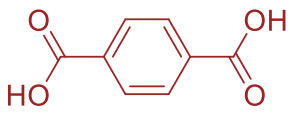

**$H_2$A1**

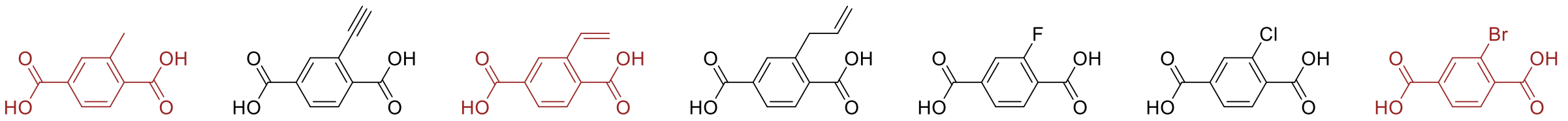

**$H_2$A2** **$H_2$A3** **$H_2$A4** **$H_2$A5** **$H_2$A6** **$H_2$A7** **$H_2$A8**

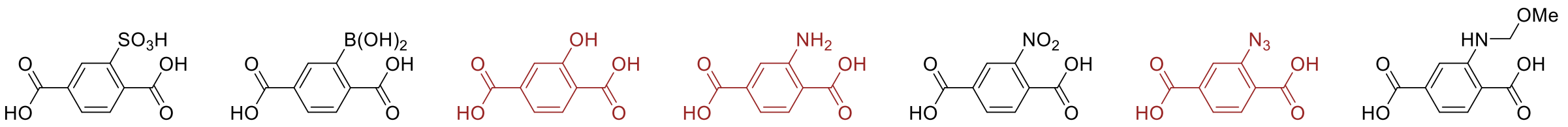

**$H_2$A9** **$H_2$A10** **$H_2$A11** **$H_2$A12** **$H_2$A13** **$H_2$A14** **$H_2$A15**

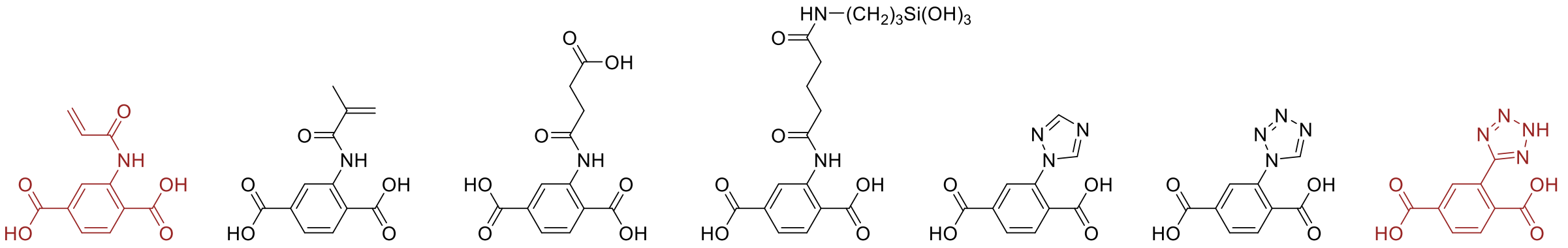

**$H_2$A16** **$H_2$A17** **$H_3$A18** **$H_2$A19** **$H_2$A20** **$H_2$A21** **$H_2$A22**

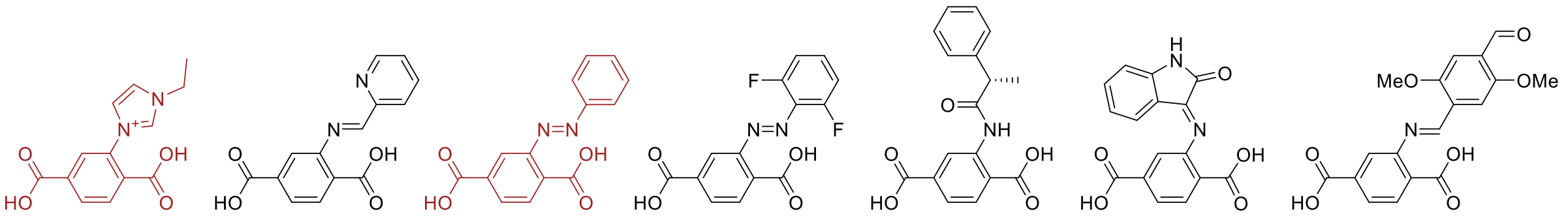

**$H_2$A23** **$H_2$A24** **$H_2$A25** **$H_2$A26** **$H_2$A27** **$H_2$A28** **$H_2$A29**

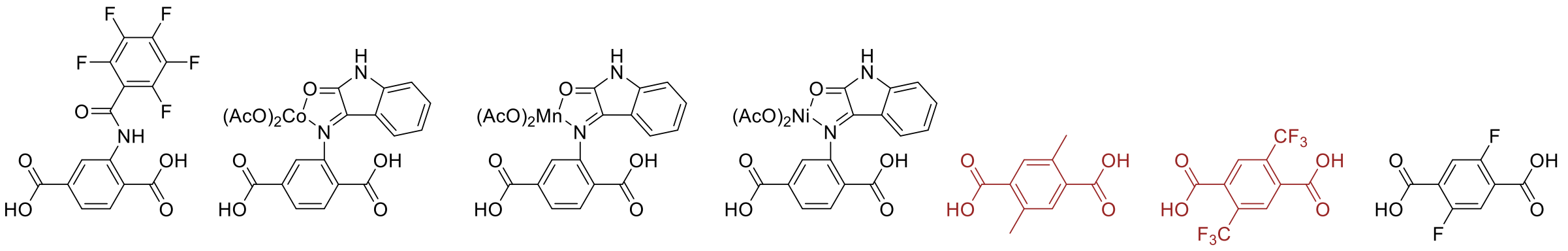

**$H_2$A30** **$H_2$A31** **$H_2$A32** **$H_2$A33** **$H_2$A34** **$H_2$A35** **$H_2$A36**

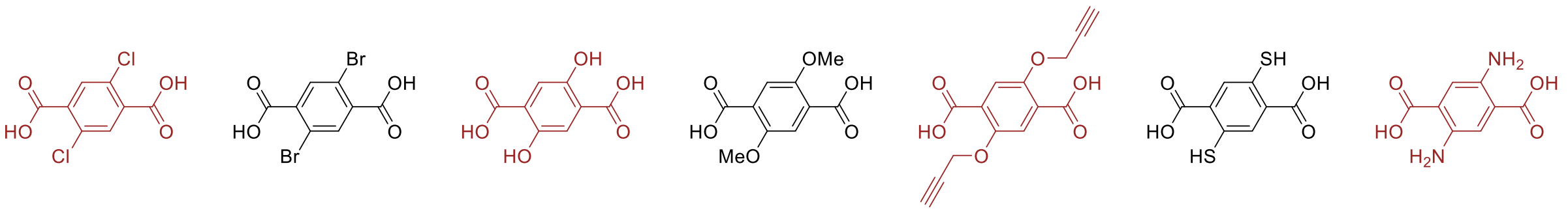

**$H_2$A37** **$H_2$A38** **$H_2$A39** **$H_2$A40** **$H_2$A41** **$H_2$A42** **$H_2$A43**

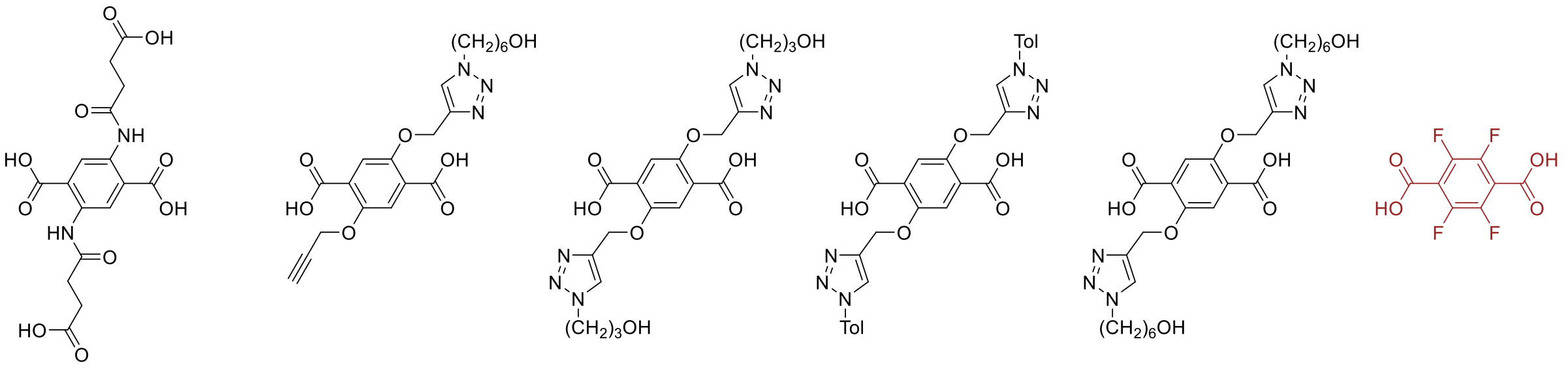

**$H_4$A44** **$H_2$A45** **$H_2$A46** **$H_2$A47** **$H_2$A48** **$H_2$A49**

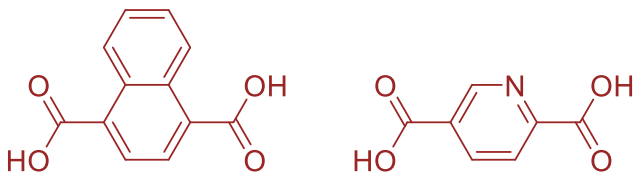

**$H_2$A50** **$H_2$A51**

**Figure 3.** Linear ditopic linkers based on benzene-1,4-dicarboxylic acid ($H_2BDC$) used to construct Zr-MOPs. Linkers in red have a reported Zr-MOP crystallographic information file (CIF). **$H_2$A1:** benzene-1,4-dicarboxylic acid; **$H_2$A2:** 2-methylbenzene-1,4-dicarboxylic acid; **$H_2$A3:** 2-ethynylbenzene-1,4-dicarboxylic acid; **$H_2$A4:** 2-ethenylbenzene-1,4-dicarboxylic acid; **$H_2$A5:** 2-(prop-2-en-1-yl)benzene-1,4-dicarboxylic acid; **$H_2$A6:** 2-fluorobenzene-1,4-dicarboxylic acid; **$H_2$A7:** 2-chlorobenzene-1,4-dicarboxylic acid; **$H_2$A8:** 2-bromobenzene-1,4-dicarboxylic acid; **$H_2$A9:** 2-sulfobenzene-1,4-dicarboxylic acid; **$H_2$A10:** 2-boronobenzene-1,4-dicarboxylic acid; **$H_2$A11:** 2-hydroxybenzene-1,4-dicarboxylic acid; **$H_2$A12:** 2-aminobenzene-1,4-dicarboxylic acid; **$H_2$A13:** 2-nitrobenzene-1,4-dicarboxylic acid; **$H_2$A14:** 2-azidobenzene-1,4-dicarboxylic acid; **$H_2$A15:** 2-[(methoxymethyl)amino]benzene-1,4-dicarboxylic acid; **$H_2$A16:** 2-((prop-2-enoyl)amino)benzene-1,4-dicarboxylic acid; **$H_2$A17:** 2-(2-methylprop-2-enamido)benzene-1,4-dicarboxylic acid; **$H_3$A18:** 2-(3-carboxypropanamido)benzene-1,4-dicarboxylic acid; **$H_2$A19:** 2-(5-oxo-5-{[3-(trihydroxysilyl)propyl]amino}pentanamido)benzene-1,4-dicarboxylic acid; **$H_2$A20:** 2-(1*H*-1,2,4-triazol-1-yl)benzene-1,4-dicarboxylic acid; **$H_2$A21:** 2-(1*H*-tetrazol-1-yl)benzene-1,4-dicarboxylic acid; **$H_2$A22:** 2-(2*H*-tetrazol-5-yl)benzene-1,4-dicarboxylic acid; **$H_2$A23:** 2-(1-ethyl-1*H*-imidazol-3-ium-3-yl)benzene-1,4-dicarboxylic acid; **$H_2$A24:** 2-[(pyridin-2-ylmethylidene)amino]benzene-1,4-dicarboxylic acid; **$H_2$A25:** 2-(phenyldiazenyl)benzene-1,4-dicarboxylic acid; **$H_2$A26:** 2-[(2,6-difluorophenyl)diazenyl]benzene-1,4-dicarboxylic acid; **$H_2$A27:** 2-[(2*S*)-2-phenylpropanamido]benzene-1,4-dicarboxylic acid; **$H_2$A28:** 2-[(2-oxo-1,2-dihydro-3*H*-indol-3-ylidene)amino]benzene-1,4-dicarboxylic acid; **$H_2$A29:** 2-{[(4-formyl-2,5-dimethoxyphenyl)methylidene]amino}benzene-1,4-dicarboxylic acid; **$H_2$A30:** 2-(2,3,4,5,6-pentafluorobenzamido)benzene-1,4-dicarboxylic acid; **$H_2$A31:** bis(acetato-κ*O*){3-[(2,5-dicarboxyphenyl)imino-κ*N*]-1,3-dihydro-2*H*-indol-2-one-κ*O*}cobalt(II); **$H_2$A32:** bis(acetato-κ*O*){3-[(2,5-dicarboxyphenyl)imino-κ*N*]-1,3-dihydro-2*H*-indol-2-one-κ*O*}manganese(II); **$H_2$A33:** bis(acetato-κ*O*){3-[(2,5-dicarboxyphenyl)imino-κ*N*]-1,3-dihydro-2*H*-indol-2-one-κ*O*}nickel(II); **$H_2$A34:** 2,5-dimethylbenzene-1,4-dicarboxylic acid; **$H_2$A35:** 2,5-bis(trifluoromethyl)benzene-1,4-dicarboxylic acid; **$H_2$A36:** 2,5-difluorobenzene-1,4-dicarboxylic acid; **$H_2$A37:** 2,5-dichlorobenzene-1,4-dicarboxylic acid; **$H_2$A38:** 2,5-dibromobenzene-1,4-dicarboxylic acid; **$H_2$A39:** 2,5-dihydroxybenzene-1,4-dicarboxylic acid; **$H_2$A40:** 2,5-dimethoxybenzene-1,4-dicarboxylic acid; **$H_2$A41:** 2,5-bis((prop-2-yn-1-yl)oxy)benzene-1,4-dicarboxylic acid; **$H_2$A42:** 2,5-disulfanylbenzene-1,4-dicarboxylic acid; **$H_2$A43:** 2,5-diaminobenzene-1,4-dicarboxylic acid; **$H_4$A44:** 2,5-bis(3-carboxypropanamido)benzene-1,4-dicarboxylic acid; **$H_2$A45:** 2-{[1-(6-hydroxyhexyl)-1*H*-1,2,3-triazol-4-yl]methoxy}-5-(prop-2-yn-1-yloxy)benzene-1,4-dicarboxylic acid; **$H_2$A46:** 2,5-bis{[1-(3-hydroxypropyl)-1*H*-1,2,3-triazol-4-yl]methoxy}benzene-1,4-dicarboxylic acid; **$H_2$A47:** 2,5-bis{[1-(4-methylphenyl)-1*H*-1,2,3-triazol-4-yl]methoxy}benzene-1,4-dicarboxylic acid; **$H_2$A48:** 2,5-bis{[1-(6-hydroxyhexyl)-1*H*-1,2,3-triazol-4-yl]methoxy}benzene-1,4-dicarboxylic acid; **$H_2$A49:** 2,3,5,6-tetrafluorobenzene-1,4-dicarboxylic acid; **$H_2$A50:** naphthalene-1,4-dicarboxylic acid; **$H_2$A51:** pyridine-2,5-dicarboxylic acid.

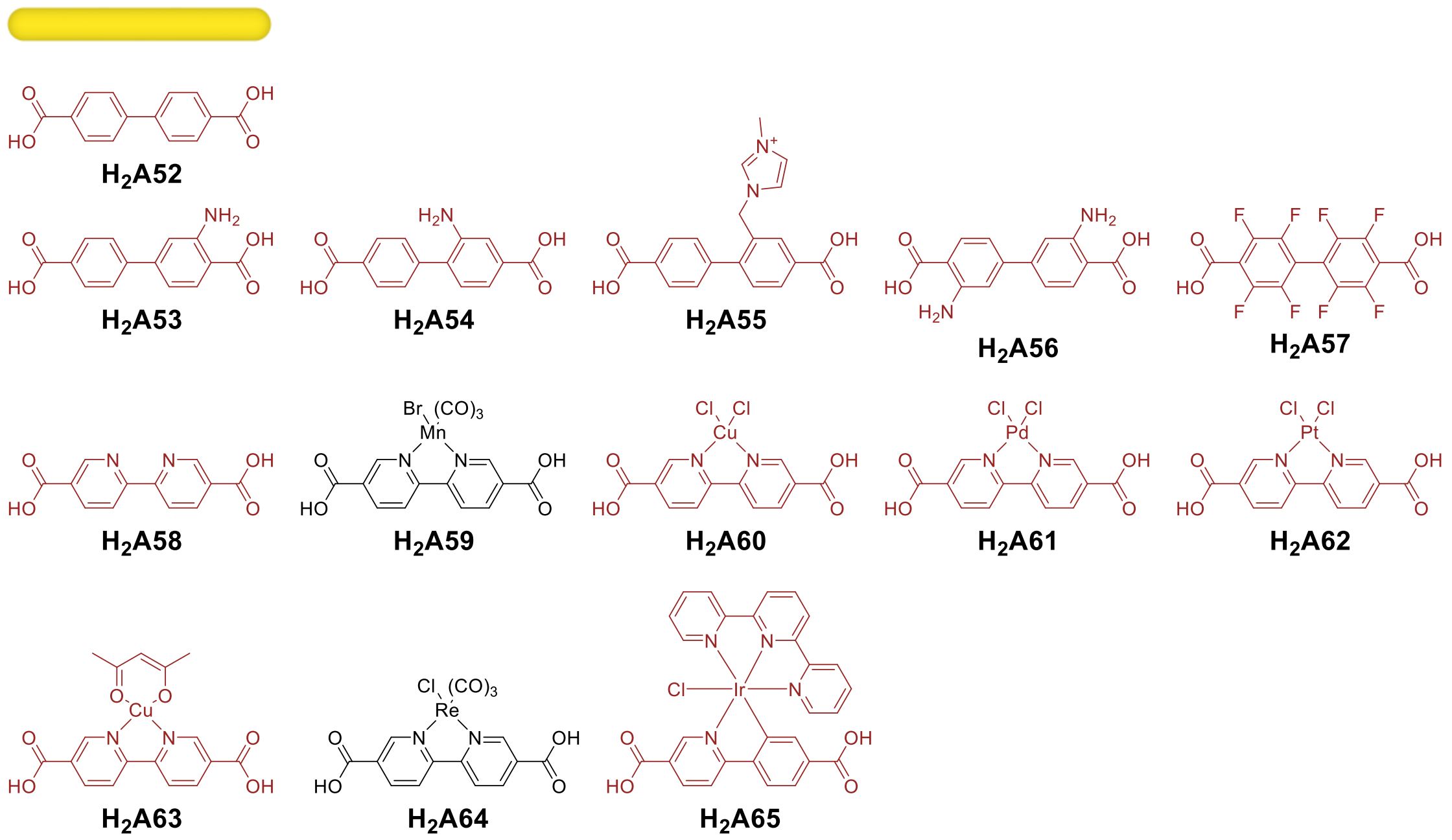

**Figure 4.** Linear ditopic linkers based on [1,1′-biphenyl]-4,4′-dicarboxylic acid (H$_2$BPDC), including [2,2′-bipyridine]-5,5′-dicarboxylic acid analogs, used to construct Zr-MOPs. Linkers in red have a reported Zr-MOP CIF. **H$_2$A52:** [1,1′-biphenyl]-4,4′-dicarboxylic acid; **H$_2$A53:** 3-amino[1,1′-biphenyl]-4,4′-dicarboxylic acid; **H$_2$A54:** 2-amino[1,1′-biphenyl]-4,4′-dicarboxylic acid; **H$_2$A55:** 2-[(3-methyl-1*H*-imidazol-3-ium-1-yl)methyl][1,1′-biphenyl]-4,4′-dicarboxylic acid; **H$_2$A56:** 3,3′-diamino[1,1′-biphenyl]-4,4′-dicarboxylic acid; **H$_2$A57:** 2,2′,3,3′,5,5′,6,6′-octafluoro[1,1′-biphenyl]-4,4′-dicarboxylic acid; **H$_2$A58:** [2,2′-bipyridine]-5,5′-dicarboxylic acid; **H$_2$A59:** bromidotricarbonyl(5,5′-dicarboxy-2,2′-bipyridine-κ$^2$*N*,*N*′)manganese(I); **H$_2$A60:** dichlorido(5,5′-dicarboxy-2,2′-bipyridine-κ$^2$*N*,*N*′)copper(II); **H$_2$A61:** dichlorido(5,5′-dicarboxy-2,2′-bipyridine-κ$^2$*N*,*N*′)palladium(II); **H$_2$A62:** dichlorido(5,5′-dicarboxy-2,2′-bipyridine-κ$^2$*N*,*N*′)platinum(II); **H$_2$A63:** (5,5′-dicarboxy-2,2′-bipyridine-κ$^2$*N*,*N*′)(pentane-2,4-dionato-κ$^2$*O*,*O*′)copper(II); **H$_2$A64:** *fac*-tricarbonylchlorido(5,5′-dicarboxy-2,2′-bipyridine-κ$^2$*N*,*N*′)rhenium(I); **H$_2$A65:** chlorido[5-carboxy-2-(5-carboxypyridin-2-yl-κ*N*)phenyl-κ*C*$^1$](2,2′:6′,2″-terpyridine-κ$^3$*N*,*N*′,*N*″)iridium(III).

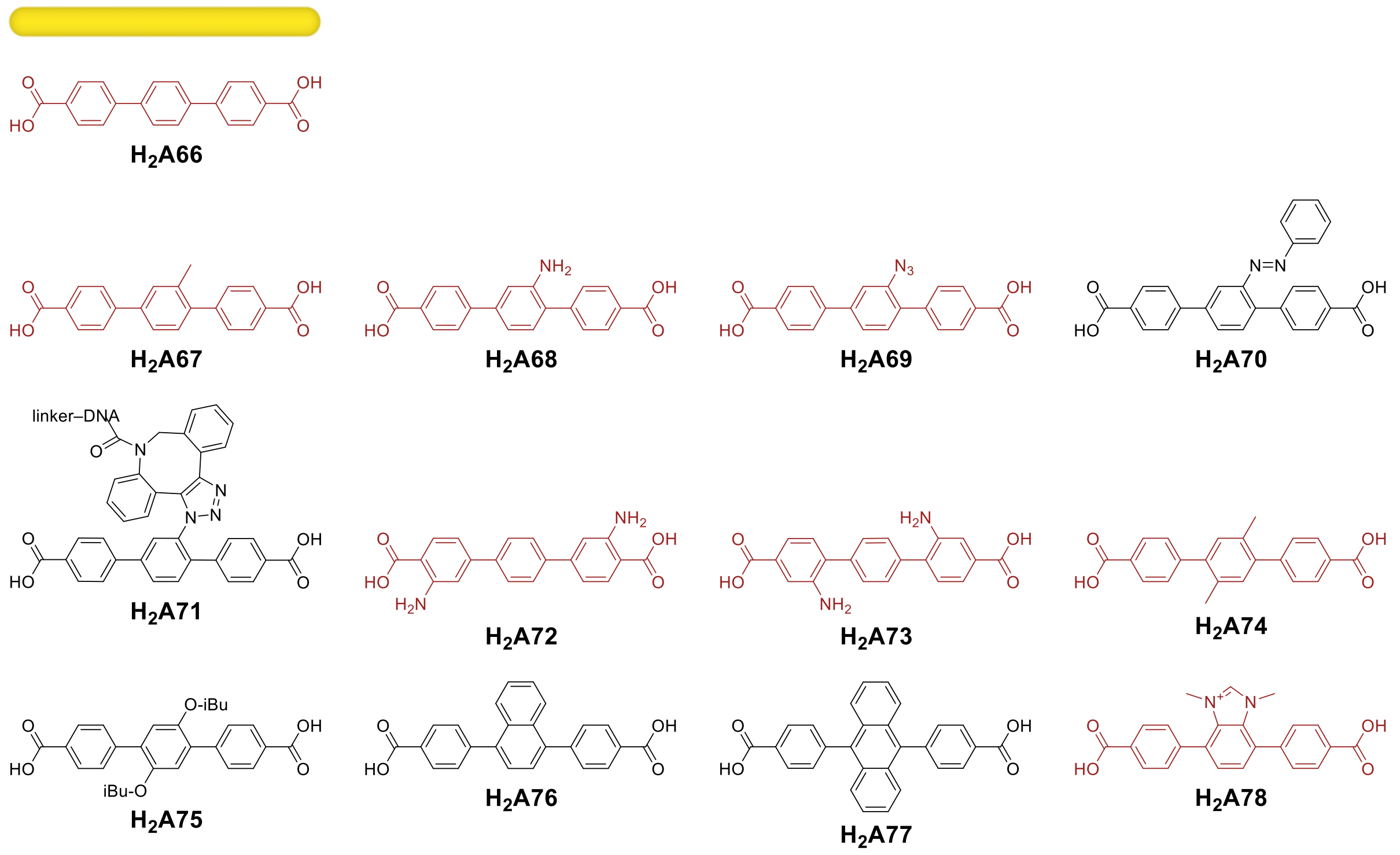

**Figure 5.** Linear ditopic linkers based on [1,1′:4′,1″-terphenyl]-4,4″-dicarboxylic acid ($H_2$TPDC), including analogs with a fused or heteroaromatic central ring, used to construct Zr-MOPs. Linkers in red have a reported Zr-MOP CIF. **H$_2$A66:** [1,1′:4′,1″-terphenyl]-4,4″-dicarboxylic acid; **H$_2$A67:** 2′-methyl[1,1′:4′,1″-terphenyl]-4,4″-dicarboxylic acid; **H$_2$A68:** 2′-amino[1,1′:4′,1″-terphenyl]-4,4″-dicarboxylic acid; **H$_2$A69:** 2′-azido[1,1′:4′,1″-terphenyl]-4,4″-dicarboxylic acid; **H$_2$A70:** 2′-(phenyldiazenyl)[1,1′:4′,1″-terphenyl]-4,4″-dicarboxylic acid; **H$_2$A71:** 2′-[DBCO–linker–DNA-triazolyl][1,1′:4′,1″-terphenyl]-4,4″-dicarboxylic acid; **H$_2$A72:** 3,3″-diamino[1,1′:4′,1″-terphenyl]-4,4″-dicarboxylic acid; **H$_2$A73:** 2,2″-diamino[1,1′:4′,1″-terphenyl]-4,4″-dicarboxylic acid; **H$_2$A74:** 2′,5′-dimethyl[1,1′:4′,1″-terphenyl]-4,4″-dicarboxylic acid; **H$_2$A75:** 2′,5′-bis(2-methylpropoxy)[1,1′:4′,1″-terphenyl]-4,4″-dicarboxylic acid; **H$_2$A76:** 4,4′-(naphthalene-1,4-diyl)dibenzoic acid; **H$_2$A77:** 4,4′-(anthracene-9,10-diyl)dibenzoic acid; **H$_2$A78:** 4,4′-(1,3-dimethyl-1*H*-benzimidazol-3-ium-4,7-diyl)dibenzoic acid.

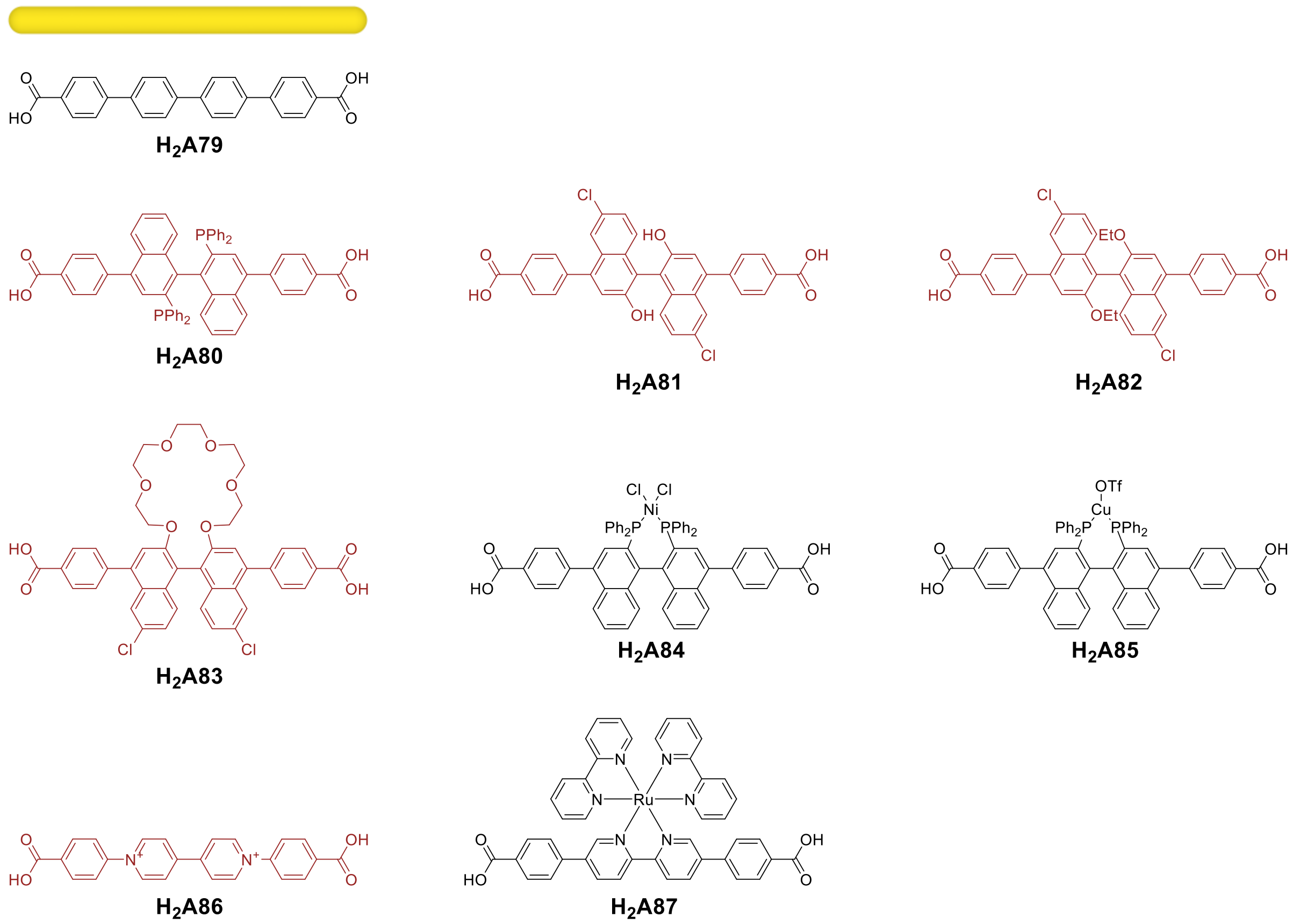

**Figure 6.** Extended linear ditopic linkers based on [1,1′:4′,1″:4″,1‴-quaterphenyl]-4,4‴-dicarboxylic acid ($H_2$QPDC) and its 1,1′-binaphthyl- and bipyridine-cored analogs used to construct Zr-MOPs. Linkers in red have a reported Zr-MOP CIF. **$H_2$A79:** [1,1′:4′,1″:4″,1‴-quaterphenyl]-4,4‴-dicarboxylic acid; **$H_2$A80:** 4,4′-[2,2′-bis(diphenylphosphanyl)[1,1′-binaphthalene]-4,4′-diyl]dibenzoic acid; **$H_2$A81:** 4,4′-(6,6′-dichloro-2,2′-dihydroxy-1,1′-binaphthalene-4,4′-diyl)dibenzoic acid; **$H_2$A82:** 4,4′-(6,6′-dichloro-2,2′-diethoxy-1,1′-binaphthalene-4,4′-diyl)dibenzoic acid; **$H_2$A83:** 4,4′-(3,26-dichloro-8,9,11,12,14,15,17,18,20,21-decahydrodinaphtho[2,1-*q*:1′,2′-*s*][1,4,7,10,13,16]hexaoxacycloicosine-5,24-diyl)dibenzoic acid; **$H_2$A84:** [4,4′-bis(4-carboxyphenyl)-2,2′-bis(diphenylphosphanyl)-1,1′-binaphthalene-κ$^2$*P*,*P*′]dichloridonickel(II); **$H_2$A85:** [4,4′-bis(4-carboxyphenyl)-2,2′-bis(diphenylphosphanyl)-1,1′-binaphthalene-κ$^2$*P*,*P*′](trifluoromethanesulfonato-κ*O*)copper(I); **$H_2$A86:** 1,1′-bis(4-carboxyphenyl)-4,4′-bipyridinium; **$H_2$A87:** bis(2,2′-bipyridine-κ$^2$*N*,*N*′)[5,5′-bis(4-carboxyphenyl)-2,2′-bipyridine-κ$^2$*N*,*N*′]ruthenium(II).

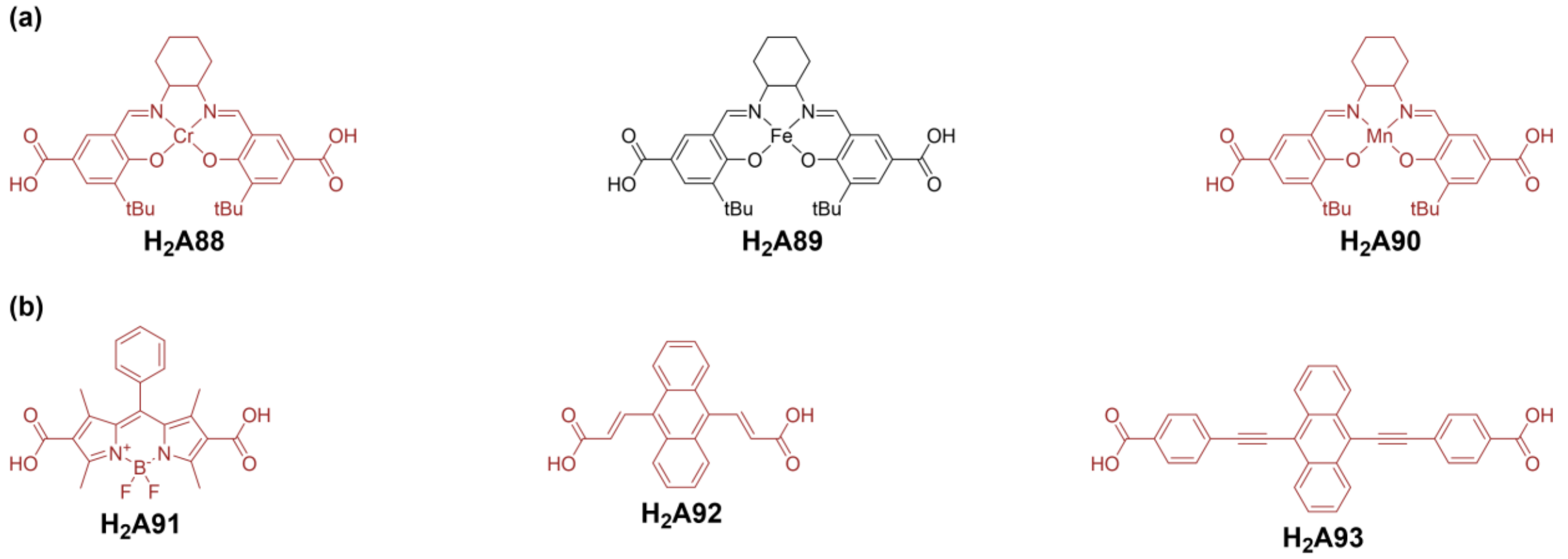

**Figure 7.** Miscellaneous linear ditopic organic linkers used to construct Zr-MOPs: (a) with metal and (b) without metal. Linkers in red have a reported Zr-MOP CIF. **$H_2$A88:** {2,2′-[cyclohexane-1,2-diylbis(azanylylidenemethanylylidene)]bis(6-*tert*-butyl-4-carboxyphenolato)-$\kappa^4$*N*,*N*′,*O*,*O*′}chromium(III); **$H_2$A89:** {2,2′-[cyclohexane-1,2-diylbis(azanylylidenemethanylylidene)]bis(6-*tert*-butyl-4-carboxyphenolato)-$\kappa^4$*N*,*N*′,*O*,*O*′}iron(III); **$H_2$A90:** {2,2′-[cyclohexane-1,2-diylbis(azanylylidenemethanylylidene)]bis(6-*tert*-butyl-4-carboxyphenolato)-$\kappa^4$*N*,*N*′,*O*,*O*′}manganese(III); **$H_2$A91:** 5,5-difluoro-1,3,7,9-tetramethyl-10-phenyl-5*H*-4$\lambda^4$,5$\lambda^4$-dipyrrolo[1,2-*c*:2′,1′-*f*][1,3,2]diazaborinine-2,8-dicarboxylic acid; **$H_2$A92:** 3,3′-(anthracene-9,10-diyl)di(prop-2-enoic acid); **$H_2$A93:** 4,4′-[anthracene-9,10-diyldi(ethyne-2,1-diyl)]dibenzoic acid.

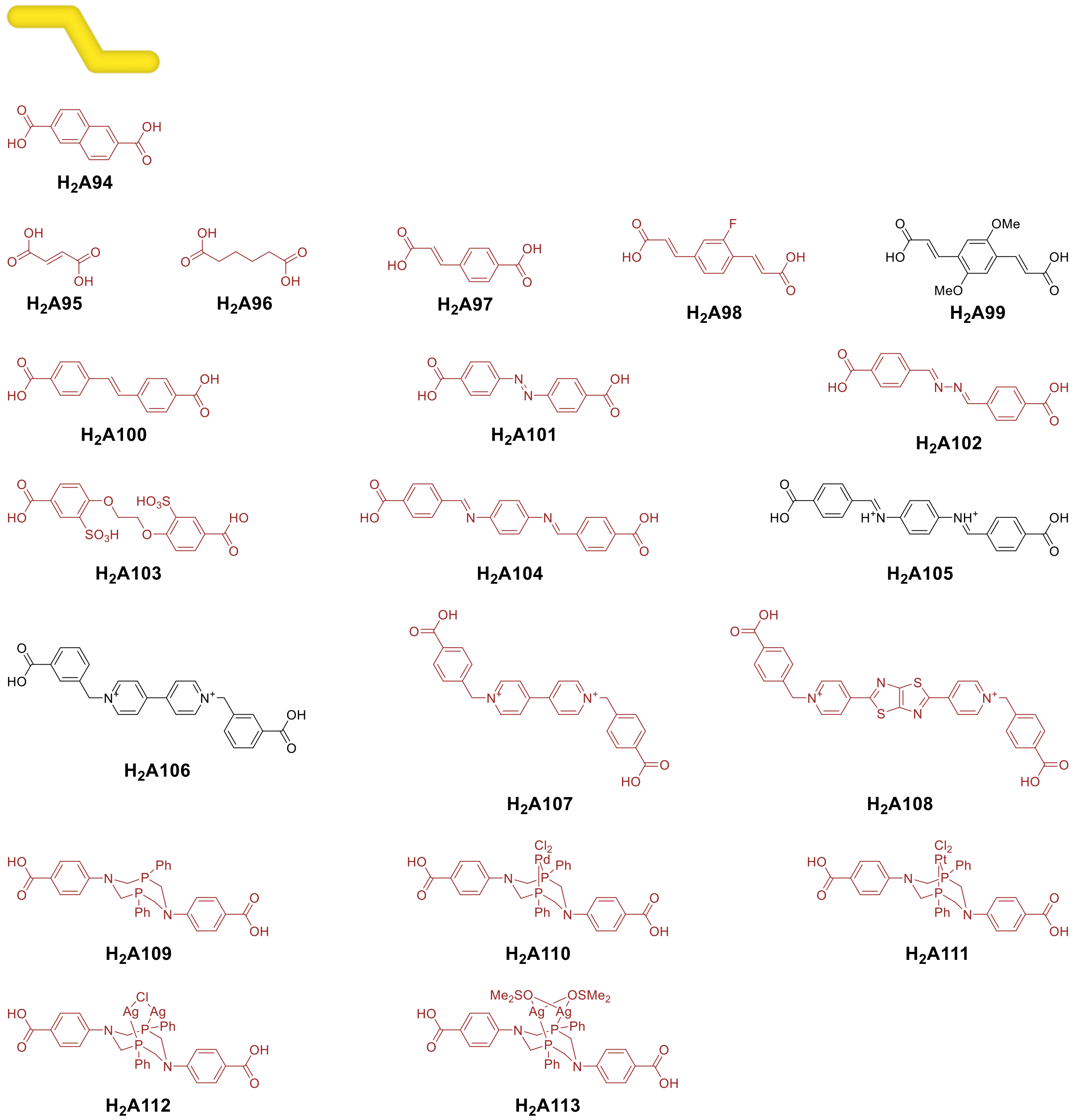

**Figure 8.** Zigzag-shaped ditopic organic linkers used to construct Zr-MOPs. Linkers in red have a reported Zr-MOP CIF. **H$_2$A94:** naphthalene-2,6-dicarboxylic acid; **H$_2$A95:** (2*E*)-but-2-enedioic acid; **H$_2$A96:** hexanedioic acid; **H$_2$A97:** 4-(2-carboxyethenyl)benzoic acid; **H$_2$A98:** (2*E*,2′*E*)-3,3′-(2-fluoro-1,4-phenylene)di(prop-2-enoic acid); **H$_2$A99:** (2*E*,2′*E*)-3,3′-(2,5-dimethoxy-1,4-phenylene)di(prop-2-enoic acid); **H$_2$A100:** 4,4′-[(*E*)-ethene-1,2-diyl]dibenzoic acid; **H$_2$A101:** 4,4′-diazenediyldibenzoic acid; **H$_2$A102:** 4,4′-(hydrazine-1,2-diylidenebis(methanylylidene))dibenzoic acid; **H$_2$A103:** 4,4′-[ethane-1,2-diylbis(oxy)]bis(3-sulfobenzoic acid); **H$_2$A104:** 4,4′-[1,4-phenylenebis(azanylylidenemethanylylidene)]dibenzoic acid; **H$_2$A105:** 4,4′-[1,4-

phenylenebis(azaniumylylidenemethanylylidene)]dibenzoic acid; **$H_2$A106:** 1,1′-bis(3-carboxybenzyl)-4,4′-bipyridinium; **$H_2$A107:** 1,1′-bis(4-carboxybenzyl)-4,4′-bipyridinium; **$H_2$A108:** 4,4′-{[1,3]thiazolo[5,4-*d*][1,3]thiazole-2,5-diylbis[(pyridin-1-ium-4,1-diyl)methylene]}dibenzoic acid; **$H_2$A109:** 4,4′-(3,7-diphenyl-1,5,3,7-diazadiphosphocane-1,5-diyl)dibenzoic acid; **$H_2$A110:** dichlorido[1,5-bis(4-carboxyphenyl)-3,7-diphenyl-1,5,3,7-diazadiphosphocane-κ$^2$*P*,*P*′]palladium(II); **$H_2$A111:** dichlorido[1,5-bis(4-carboxyphenyl)-3,7-diphenyl-1,5,3,7-diazadiphosphocane-κ$^2$*P*,*P*′]platinum(II); **$H_2$A112:** μ-chlorido-[μ-1,5-bis(4-carboxyphenyl)-3,7-diphenyl-1,5,3,7-diazadiphosphocane-κ*P*:κ*P*′]disilver(I); **$H_2$A113:** bis(μ-dimethyl sulfoxide-κ*O*:κ*O*)-[μ-1,5-bis(4-carboxyphenyl)-3,7-diphenyl-1,5,3,7-diazadiphosphocane-κ*P*:κ*P*′]disilver(I).

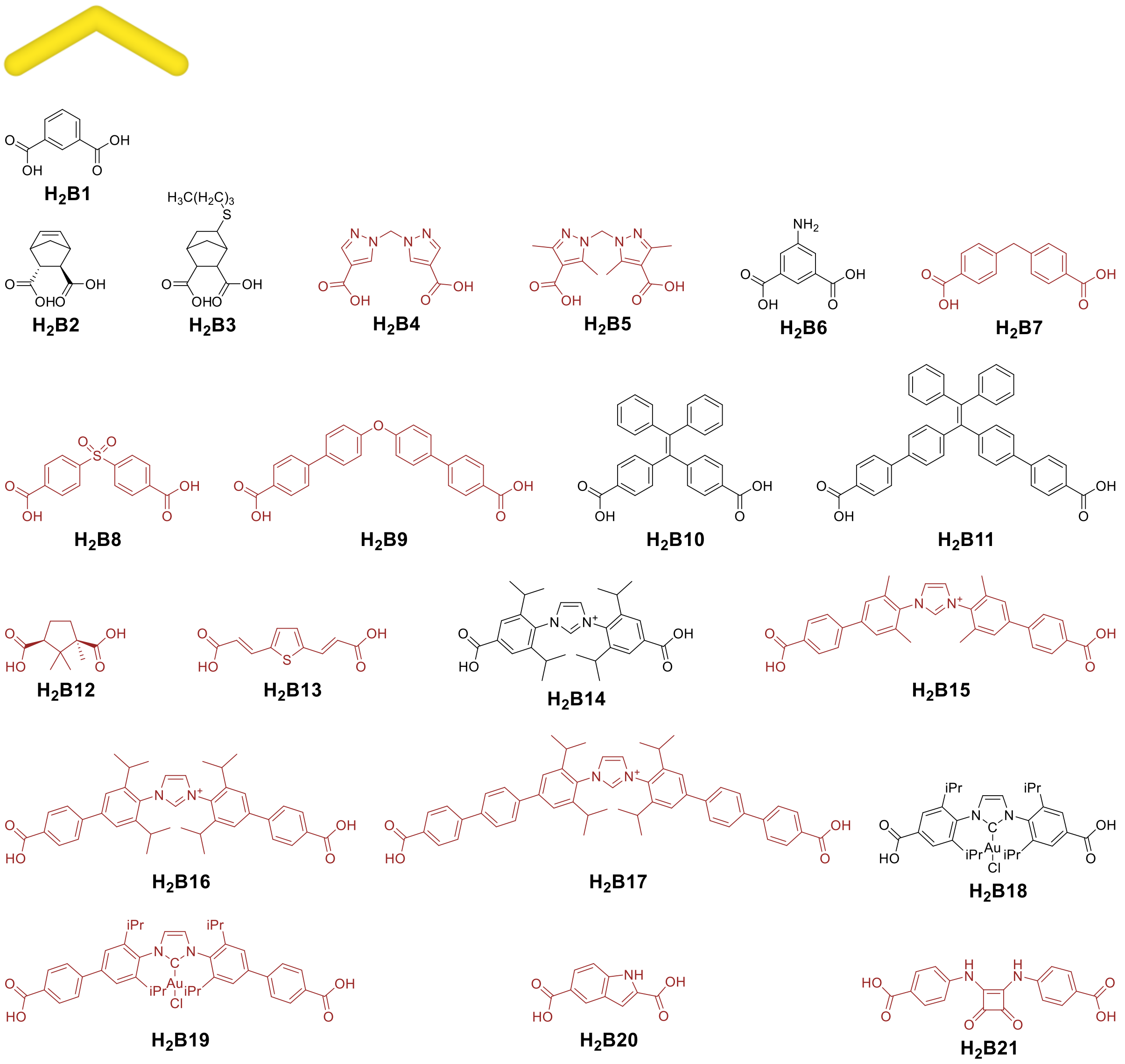

**Figure 9.** Bent ditopic organic linkers used to construct Zr-MOPs. Linkers in red have a reported Zr-MOP CIF. **$H_2$B1:** benzene-1,3-dicarboxylic acid; **$H_2$B2:** *rel*-(2*R*,3*R*)-bicyclo[2.2.1]hept-5-ene-2,3-dicarboxylic acid; **$H_2$B3:** 5-(butylthio)bicyclo[2.2.1]heptane-2,3-dicarboxylic acid; **$H_2$B4:** 1,1′-methylenebis(1*H*-pyrazole-4-carboxylic acid); **$H_2$B5:** 1,1′-methylenebis(3,5-dimethyl-1*H*-pyrazole-4-carboxylic acid); **$H_2$B6:** 5-aminobenzene-1,3-dicarboxylic acid; **$H_2$B7:** 4,4′-methylenedibenzoic acid; **$H_2$B8:** 4,4′-sulfonyldibenzoic acid; **$H_2$B9:** 4′,4‴-oxybis([1,1′-biphenyl]-4-carboxylic acid); **$H_2$B10:** 4,4′-(2,2-diphenylethene-1,1-diyl)dibenzoic acid; **$H_2$B11:** 4′,4‴-(2,2-diphenylethene-1,1-diyl)bis([1,1′-biphenyl]-4-carboxylic acid); **$H_2$B12:** (1*R*,3*S*)-1,2,2-trimethylcyclopentane-1,3-dicarboxylic acid; **$H_2$B13:** (2*E*,2′*E*)-3,3′-(thiophene-2,5-diyl)di(prop-2-enoic acid); **$H_2$B14:** 4,4′-(1*H*-imidazol-3-ium-1,3-diyl)bis[3,5-bis(propan-2-yl)benzoic acid]; **$H_2$B15:** 4′,4‴-(1*H*-imidazol-3-ium-1,3-diyl)bis(3′,5′-dimethyl[1,1′-biphenyl]-4-carboxylic acid); **$H_2$B16:** 4′,4‴-(1*H*-imidazol-3-ium-

1,3-diyl)bis[3′,5′-bis(propan-2-yl)[1,1′-biphenyl]-4-carboxylic acid]; **$H_2$B17:** 4″,4‴″-(1*H*-imidazol-3-ium-1,3-diyl)bis[3″,5″-bis(propan-2-yl)[1,1′:4′,1″-terphenyl]-4-carboxylic acid]; **$H_2$B18:** chlorido{1,3-bis[4-carboxy-2,6-bis(propan-2-yl)phenyl]imidazol-2-ylidene-κ$C^2$}gold(I); **$H_2$B19:** chlorido{1,3-bis[4′-carboxy-3,5-bis(propan-2-yl)[1,1′-biphenyl]-4-yl]imidazol-2-ylidene-κ$C^2$}gold(I); **$H_2$B20:** 1*H*-indole-2,5-dicarboxylic acid; **$H_2$B21:** 4,4′-((3,4-dioxocyclobut-1-ene-1,2-diyl)bis(azanediyl))dibenzoic acid.

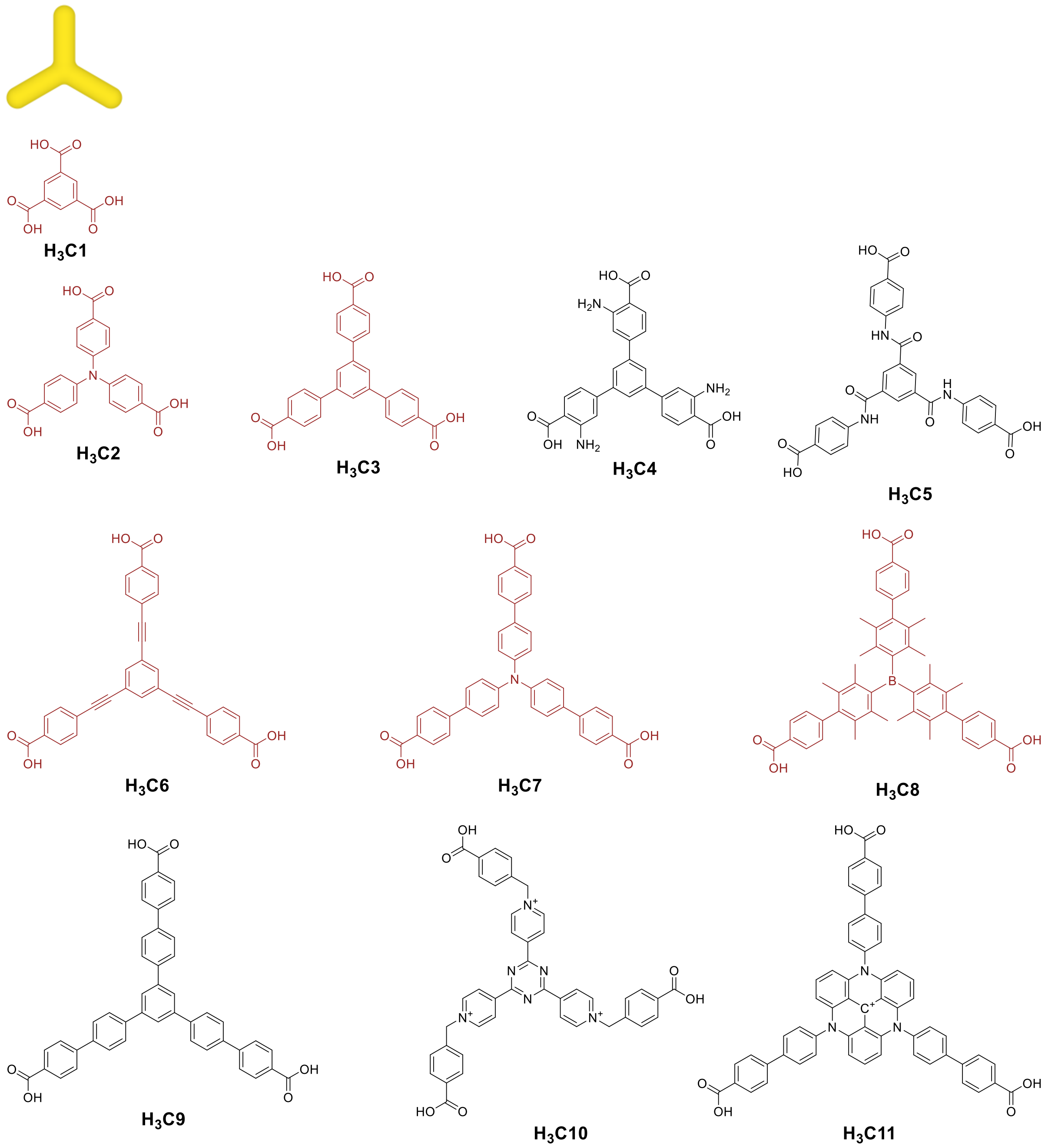

**Figure 10.** Tritopic organic linkers with three-fold symmetry as 3-c nodes to construct Zr-MOPs. Linkers in red have a reported Zr-MOP CIF. **H$_3$C1:** benzene-1,3,5-tricarboxylic acid; **H$_3$C2:** 4,4′,4″-nitrilotribenzoic acid; **H$_3$C3:** 4,4′,4″-(benzene-1,3,5-triyl)tribenzoic acid; **H$_3$C4:** 3,3″-diamino-5′-(3-amino-4-carboxyphenyl)[1,1′:3′,1″-terphenyl]-4,4″-dicarboxylic acid; **H$_3$C5:** 4,4′,4″-[1,3,5-benzenetriyltris(carbonylimino)]trisbenzoic acid; **H$_3$C6:** 4,4′,4″-(benzene-1,3,5-triyltris(ethyne-2,1-diyl))tribenzoic acid; **H$_3$C7:** 4′,4‴,4″‴-nitrilotris([1,1′-biphenyl]-4-carboxylic acid); **H$_3$C8:** 4,4′,4″-[boranetriyltris(2,3,5,6-tetramethylbenzene-4,1-diyl)]tribenzoic acid; **H$_3$C9:** 4,4′,4″-(benzene-1,3,5-triyl-tris(benzene-4,1-diyl))tribenzoic acid;

**$H_3C10$:** 1,1′,1″-tris(4-carboxybenzyl)-4,4′,4″-(1,3,5-triazine-2,4,6-triyl)tripyridinium; **$H_3C11$:** 4,8,12-tris(4′-carboxy[1,1′-biphenyl]-4-yl)-8,12-dihydro-4*H*-benzo[8,1][2,7]naphthyridino[3,4,5,6-*klmn*]acridin-12c-ylium.

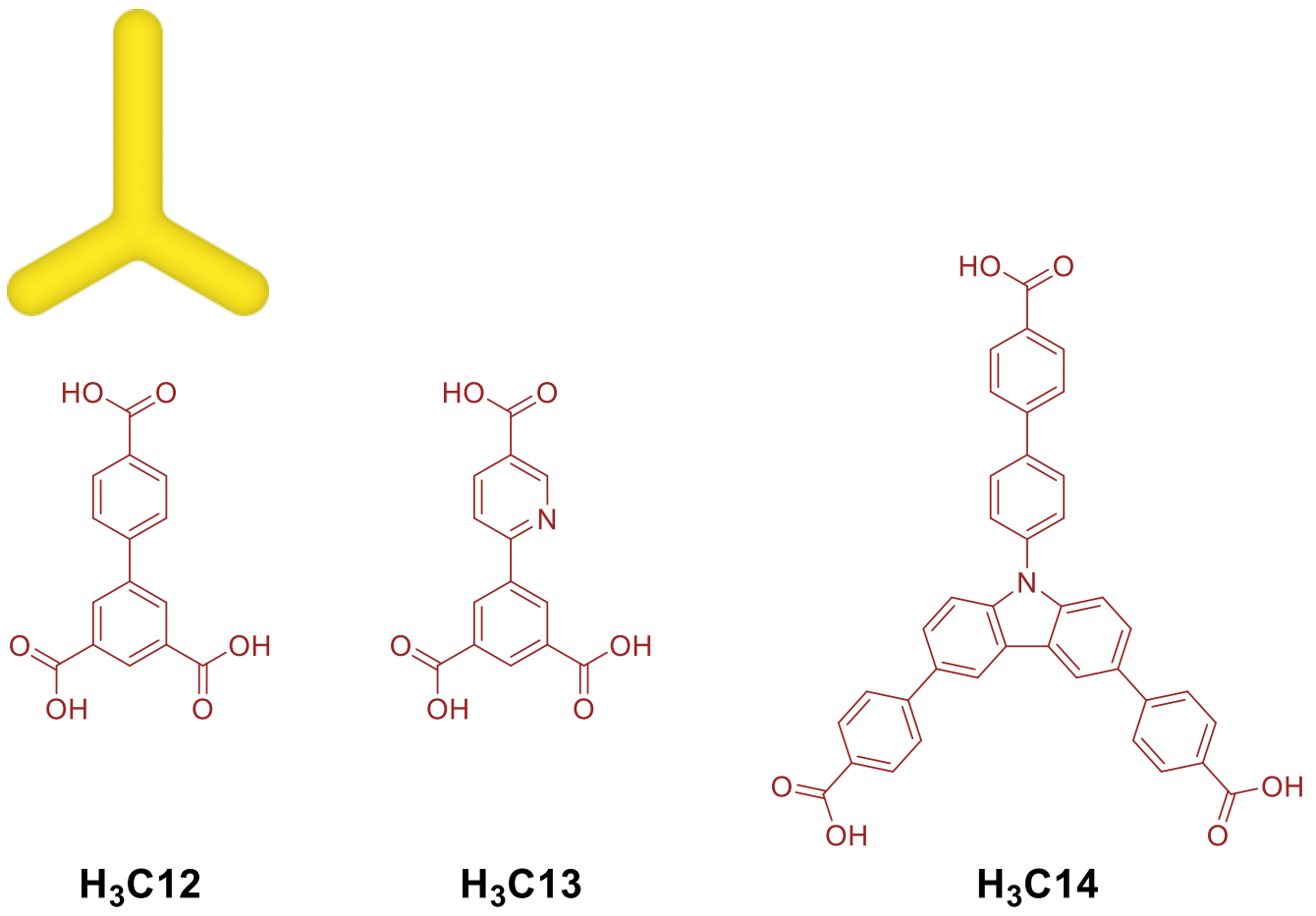

**Figure 11.** Tritopic organic linkers without three-fold symmetry as 3-c nodes to construct Zr-MOPs. Linkers in red have a reported Zr-MOP CIF. **$H_3C12$:** [1,1′-biphenyl]-3,4′,5-tricarboxylic acid; **$H_3C13$:** 5-(5-carboxypyridin-2-yl)benzene-1,3-dicarboxylic acid; **$H_3C14$:** 4′-[3,6-bis(4-carboxyphenyl)-9*H*-carbazol-9-yl][1,1′-biphenyl]-4-carboxylic acid.

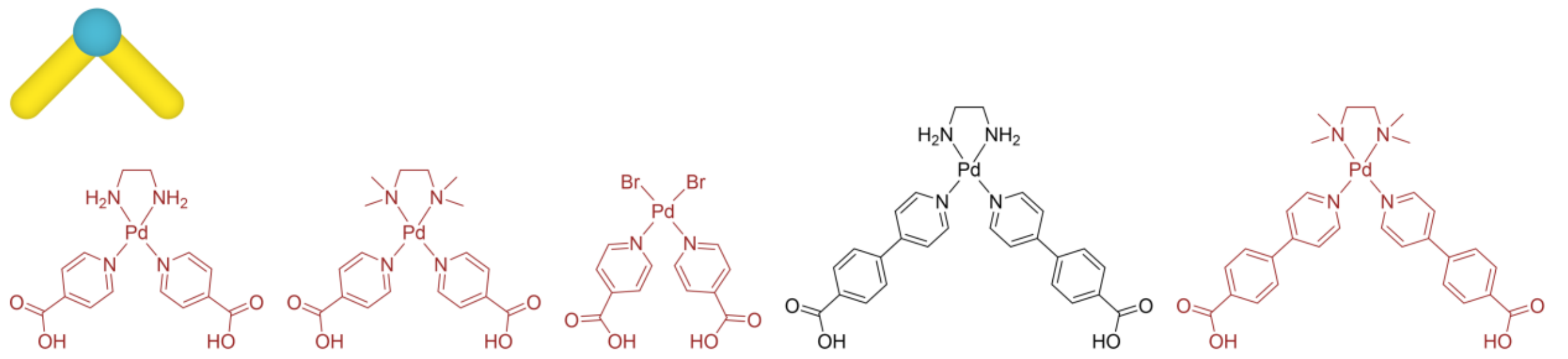

$H_2$D1 $H_2$D2 $H_2$D3 $H_2$D4 $H_2$D5

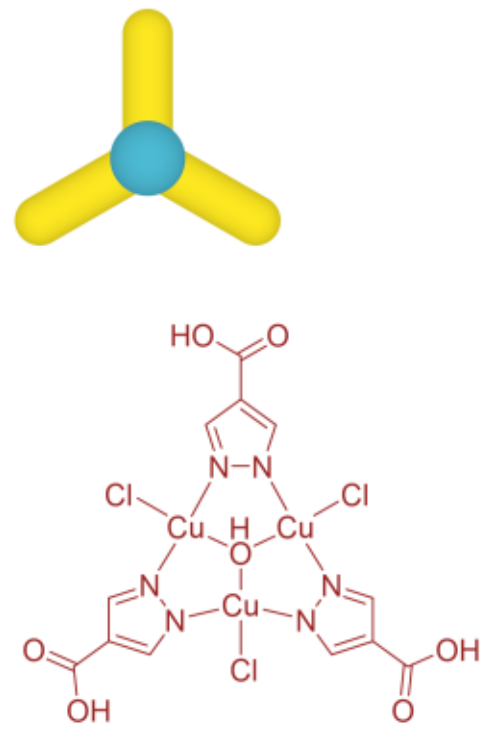

$H_3$D6

**Figure 12.** Metal-bridged ditopic and tritopic linkers used to construct Zr-MOPs. Linkers in red have a reported Zr-MOP CIF. **$H_2$D1:** bis(4-carboxypyridine-κ*N*)(ethane-1,2-diamine-κ$^2$*N*,*N*′)palladium(II); **$H_2$D2:** bis(4-carboxypyridine-κ*N*)(*N*,*N*,*N*′,*N*′-tetramethylethane-1,2-diamine-κ$^2$*N*,*N*′)palladium(II); **$H_2$D3:** dibromidobis(4-carboxypyridine-κ*N*)palladium(II); **$H_2$D4:** bis[4-(4-carboxyphenyl)pyridine-κ*N*](ethane-1,2-diamine-κ$^2$*N*,*N*′)palladium(II); **$H_2$D5:** bis[4-(4-carboxyphenyl)pyridine-κ*N*](*N*,*N*,*N*′,*N*′-tetramethylethane-1,2-diamine-κ$^2$*N*,*N*′)palladium(II); **$H_3$D6:** tris(μ-4-carboxypyrazolato-κ$N^1$:κ$N^2$)trichlorido(μ$_3$-hydroxido)tricopper(II).

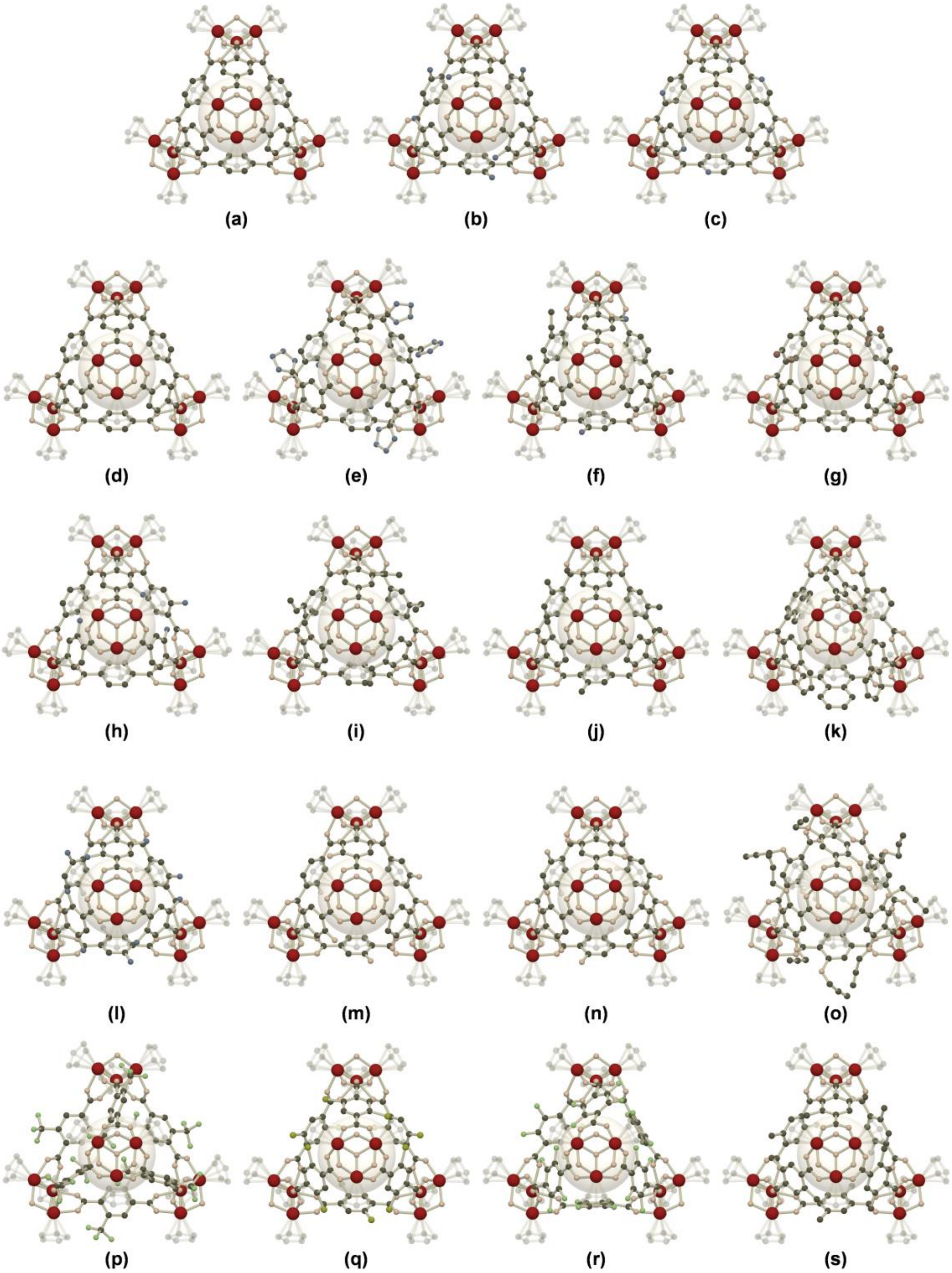

**Figure 13.** Structural illustrations of edge-directed $M_4L_6$ tetrahedra, $[Cp_3Zr_3(\mu_3\text{-O})(\mu_2\text{-OH})_3]_4(\mathbf{L})_6$, built from BDC-based linkers. (a) $[Cp_3Zr_3(\mu_3\text{-O})(\mu_2\text{-OH})_3]_4(\mathbf{A1})_6$; (b) $[Cp_3Zr_3(\mu_3\text{-O})(\mu_2\text{-OH})_3]_4(\mathbf{A16})_6$; (c) $[Cp_3Zr_3(\mu_3\text{-O})(\mu_2\text{-OH})_3]_4(\mathbf{A51})_6$; (d) $[Cp_3Zr_3(\mu_3\text{-O})(\mu_2\text{-OH})_3]_4(\mathbf{A25})_6$; (e) $[Cp_3Zr_3(\mu_3\text{-O})(\mu_2\text{-OH})_3]_4(\mathbf{A22})_6$; (f) $[Cp_3Zr_3(\mu_3\text{-O})(\mu_2\text{-OH})_3]_4(\mathbf{A4})_3(\mathbf{A12})_3$; (g) $[Cp_3Zr_3(\mu_3\text{-O})(\mu_2\text{-OH})_3]_4(\mathbf{A8})_6$; (h) $[Cp_3Zr_3(\mu_3\text{-O})(\mu_2\text{-OH})_3]_4(\mathbf{A12})_6$; (i) $[Cp_3Zr_3(\mu_3\text{-O})(\mu_2\text{-OH})_3]_4(\mathbf{A4})_6$; (j) $[Cp_3Zr_3(\mu_3\text{-O})(\mu_2\text{-OH})_3]_4(\mathbf{A2})_6$; (k) $[Cp_3Zr_3(\mu_3\text{-O})(\mu_2\text{-OH})_3]_4(\mathbf{A50})_6$; (l) $[Cp_3Zr_3(\mu_3\text{-O})(\mu_2\text{-OH})_3]_4(\mathbf{A43})_6$; (m) $[Cp_3Zr_3(\mu_3\text{-O})(\mu_2\text{-OH})_3]_4(\mathbf{A11})_6$; (n) $[Cp_3Zr_3(\mu_3\text{-O})(\mu_2\text{-OH})_3]_4(\mathbf{A39})_6$; (o) $[Cp_3Zr_3(\mu_3\text{-O})(\mu_2\text{-OH})_3]_4(\mathbf{A41})_6$; (p) $[Cp_3Zr_3(\mu_3\text{-O})(\mu_2\text{-OH})_3]_4(\mathbf{A35})_6$; (q) $[Cp_3Zr_3(\mu_3\text{-O})(\mu_2\text{-}$

OH)$_3$]$_4$(**A37**)$_6$; (r) [Cp$_3$Zr$_3$(μ$_3$-O)(μ$_2$-OH)$_3$]$_4$(**A49**)$_6$; (s) [Cp$_3$Zr$_3$(μ$_3$-O)(μ$_2$-OH)$_3$]$_4$(**A34**)$_6$. Cage structures are drawn from the atomic coordinates in the deposited CIFs after removal of crystallographic disorder; partially occupied components are distributed over equivalent sites of the cage in proportion to their occupancies, and atoms not located crystallographically are omitted.

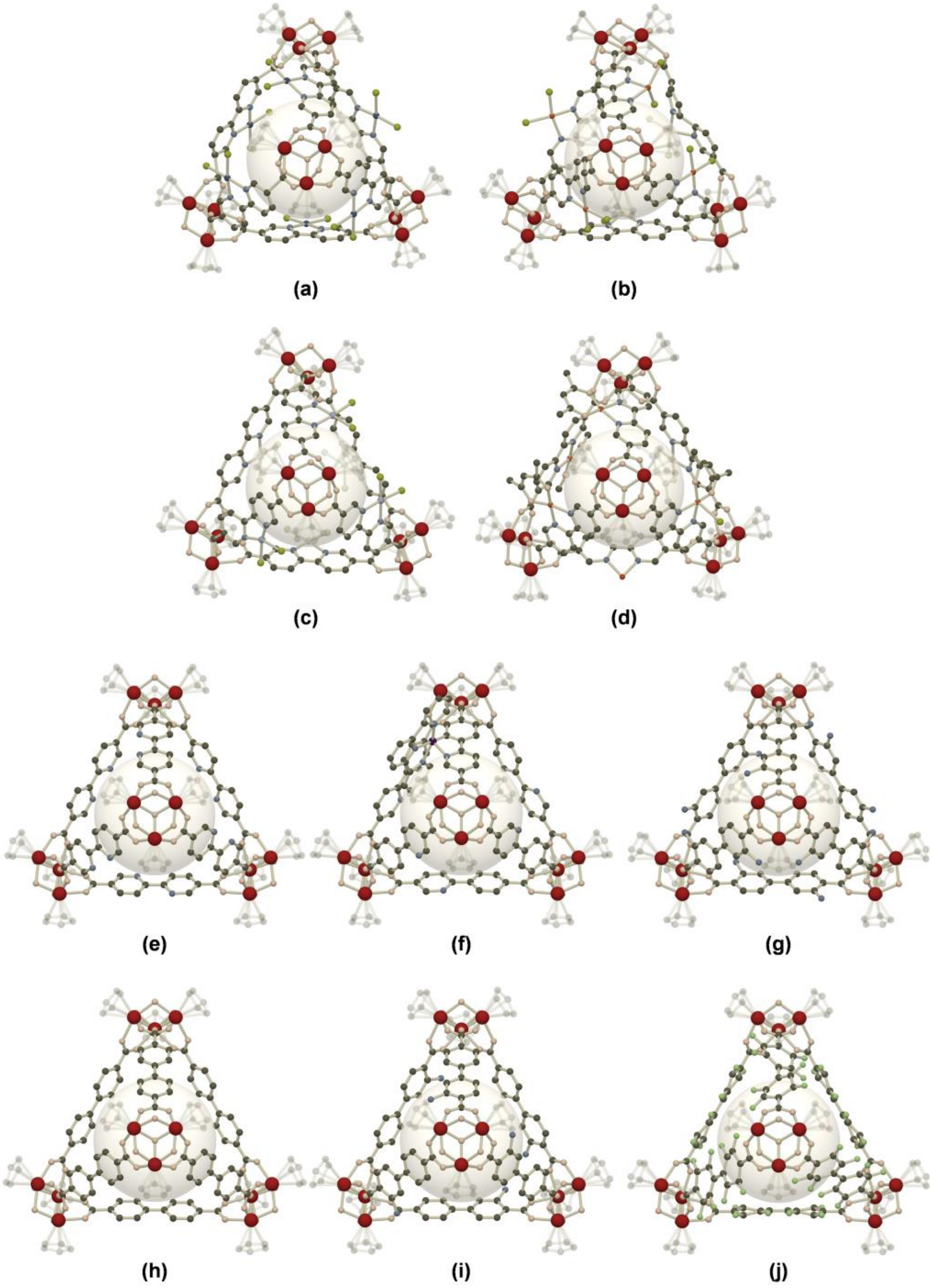

**Figure 14.** Structural illustrations of edge-directed M$_4$L$_6$ tetrahedra, [Cp$_3$Zr$_3$(μ$_3$-O)(μ$_2$-OH)$_3$]$_4$(**L**)$_6$, built from BPDC-based linkers. (a) [Cp$_3$Zr$_3$(μ$_3$-O)(μ$_2$-OH)$_3$]$_4$(**A61**)$_6$; (b) [Cp$_3$Zr$_3$(μ$_3$-O)(μ$_2$-OH)$_3$]$_4$(**A60**)$_6$; (c) [Cp$_3$Zr$_3$(μ$_3$-O)(μ$_2$-OH)$_3$]$_4$(**A62**)$_6$; (d) [Cp$_3$Zr$_3$(μ$_3$-O)(μ$_2$-OH)$_3$]$_4$(**A63**)$_6$; (e) [Cp$_3$Zr$_3$(μ$_3$-O)(μ$_2$-OH)$_3$]$_4$(**A58**)$_6$; (f) [Cp$_3$Zr$_3$(μ$_3$-O)(μ$_2$-

$OH)_3]_4$(**A54**)$_{5.5}$(**A65**)$_{0.5}$; (g) $[Cp_3Zr_3(\mu_3\text{-}O)(\mu_2\text{-}OH)_3]_4$(**A56**)$_6$; (h) $[Cp_3Zr_3(\mu_3\text{-}O)(\mu_2\text{-}OH)_3]_4$(**A52**)$_6$; (i) $[Cp_3Zr_3(\mu_3\text{-}O)(\mu_2\text{-}OH)_3]_4$(**A53**)$_6$; (j) $[Cp_3Zr_3(\mu_3\text{-}O)(\mu_2\text{-}OH)_3]_4$(**A57**)$_6$.

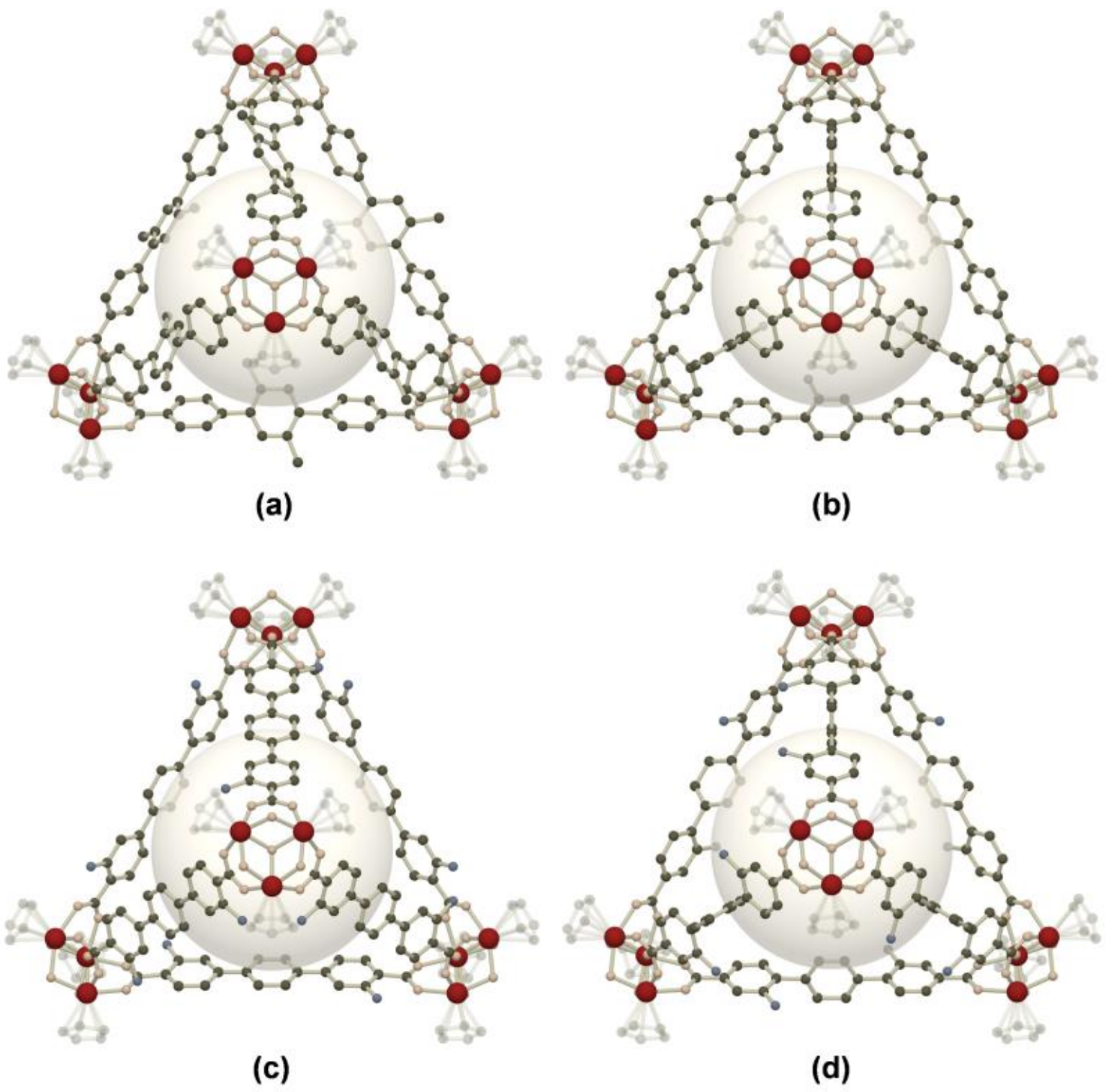

**Figure 15.** Structural illustrations of edge-directed $M_4L_6$ tetrahedra, $[Cp_3Zr_3(\mu_3\text{-}O)(\mu_2\text{-}OH)_3]_4$(**L**)$_6$, built from TPDC-based linkers. (a) $[Cp_3Zr_3(\mu_3\text{-}O)(\mu_2\text{-}OH)_3]_4$(**A74**)$_6$; (b) $[Cp_3Zr_3(\mu_3\text{-}O)(\mu_2\text{-}OH)_3]_4$(**A68**)$_6$; (c) $[Cp_3Zr_3(\mu_3\text{-}O)(\mu_2\text{-}OH)_3]_4$(**A72**)$_6$; (d) $[Cp_3Zr_3(\mu_3\text{-}O)(\mu_2\text{-}OH)_3]_4$(**A73**)$_6$.

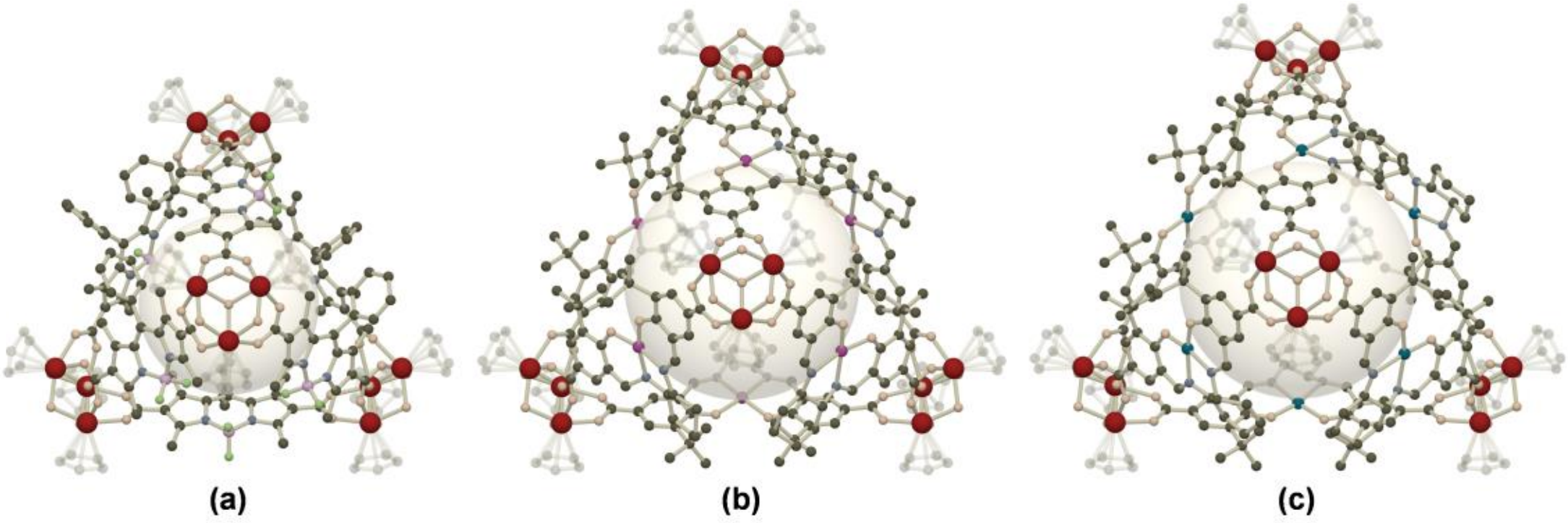

**Figure 16.** Structural illustrations of edge-directed $M_4L_6$ tetrahedra, $[Cp_3Zr_3(\mu_3\text{-}O)(\mu_2\text{-}OH)_3]_4$(**L**)$_6$, built from miscellaneous linear linkers. (a) $[Cp_3Zr_3(\mu_3\text{-}O)(\mu_2\text{-}OH)_3]_4$(**A91**)$_6$; (b) $[Cp_3Zr_3(\mu_3\text{-}O)(\mu_2\text{-}OH)_3]_4$(**A90**)$_6$; (c) $[Cp_3Zr_3(\mu_3\text{-}O)(\mu_2\text{-}OH)_3]_4$(**A88**)$_6$.

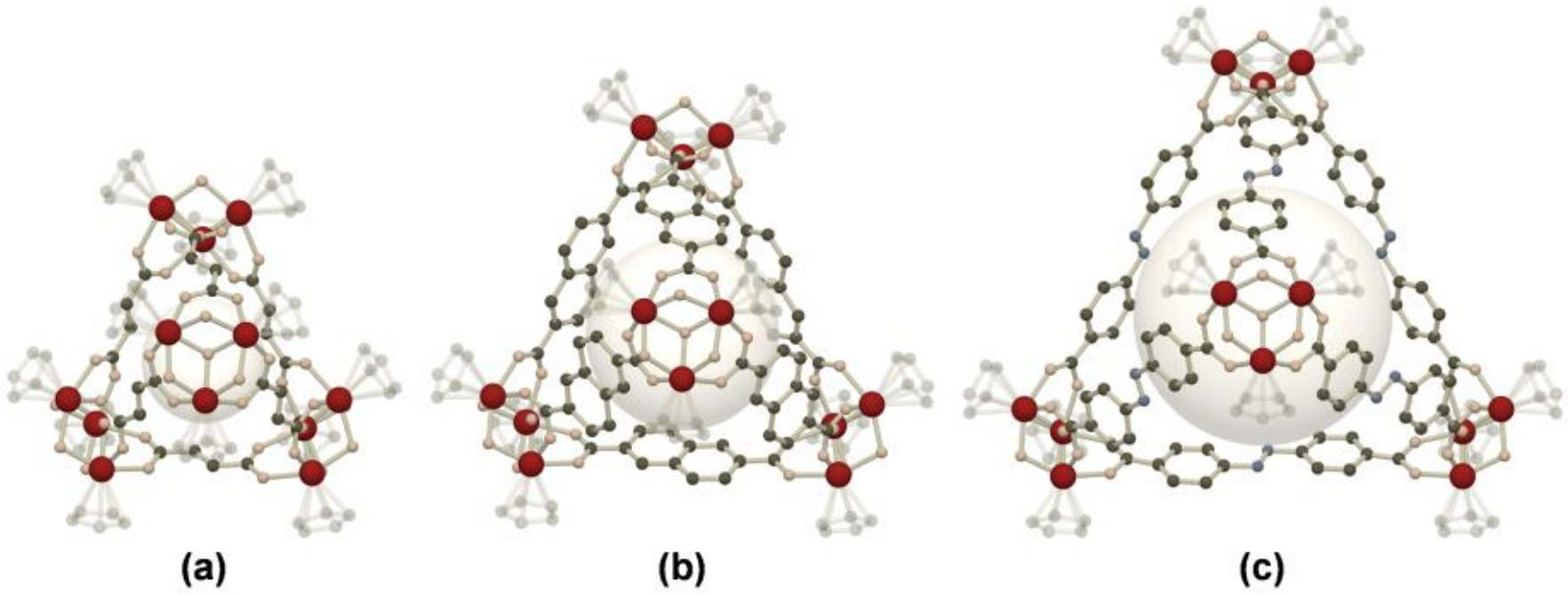

**Figure 17.** Structural illustrations of edge-directed $M_4L_6$ tetrahedra, $[Cp_3Zr_3(\mu_3\text{-}O)(\mu_2\text{-}OH)_3]_4(\mathbf{L})_6$, built from zigzag-shaped linkers. (a) $[Cp_3Zr_3(\mu_3\text{-}O)(\mu_2\text{-}OH)_3]_4(\mathbf{A95})_6$; (b) $[Cp_3Zr_3(\mu_3\text{-}O)(\mu_2\text{-}OH)_3]_4(\mathbf{A94})_6$; (c) $[Cp_3Zr_3(\mu_3\text{-}O)(\mu_2\text{-}OH)_3]_4(\mathbf{A101})_6$.

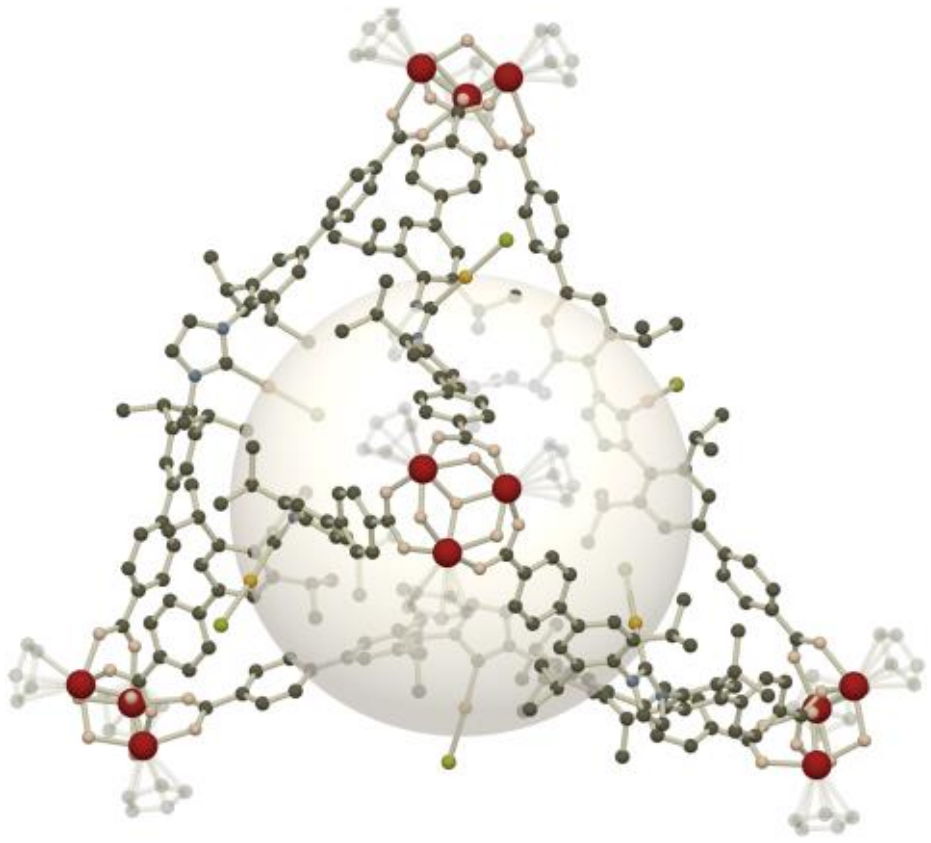

**Figure 18.** Structural illustration of an edge-directed $M_4L_6$ tetrahedron, $[Cp_3Zr_3(\mu_3\text{-}O)(\mu_2\text{-}OH)_3]_4(\mathbf{L})_6$, built from a bent linker. $[Cp_3Zr_3(\mu_3\text{-}O)(\mu_2\text{-}OH)_3]_4(\mathbf{B19})_6$.

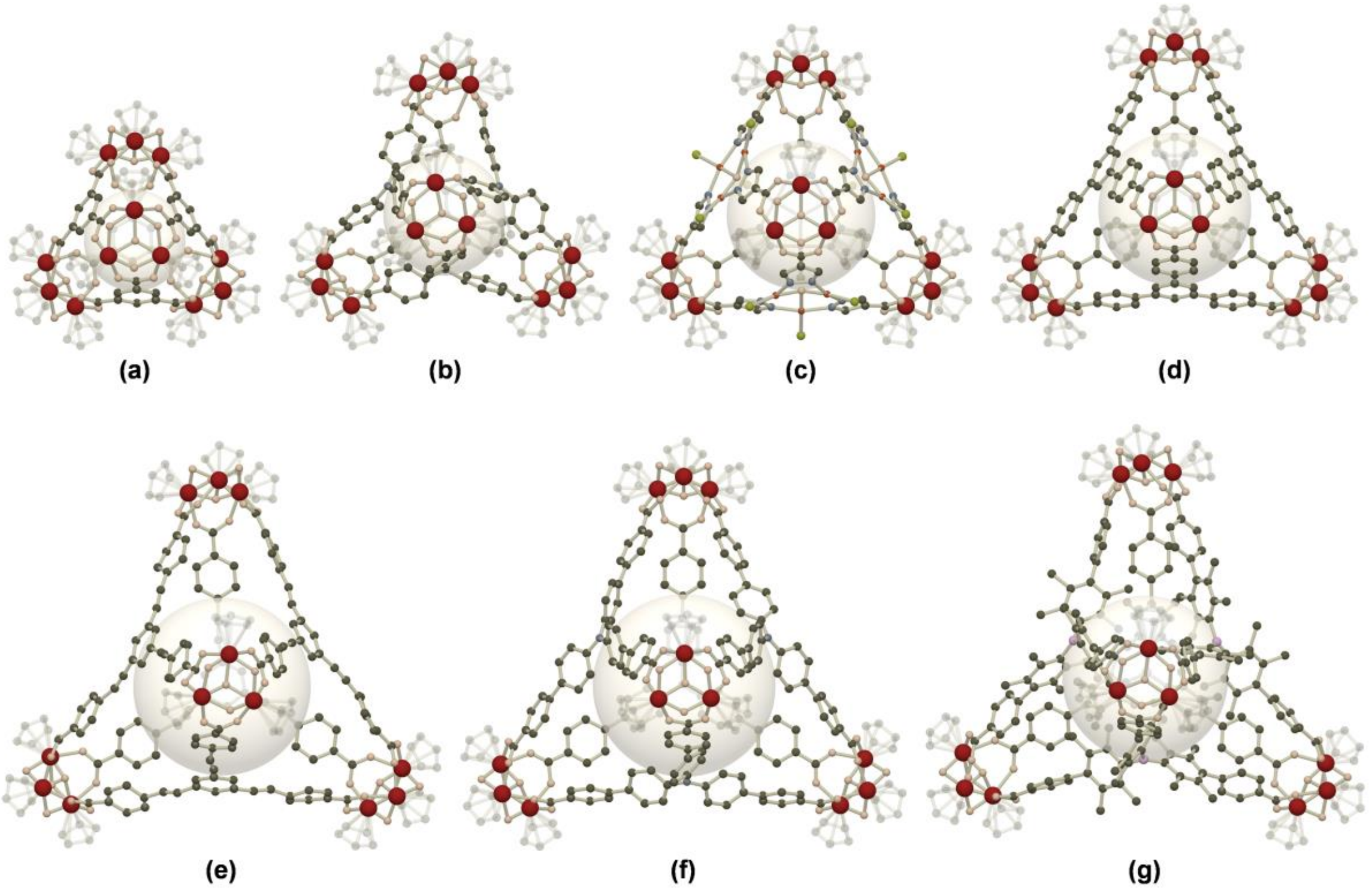

**Figure 19.** Structural illustrations of face-directed $M_4L_4$ tetrahedra, $[Cp_3Zr_3(\mu_3\text{-}O)(\mu_2\text{-}OH)_3]_4(\mathbf{L})_4$, built from tritopic linkers with three-fold symmetry. (a) $[Cp_3Zr_3(\mu_3\text{-}O)(\mu_2\text{-}OH)_3]_4(\mathbf{C1})_4$; (b) $[Cp_3Zr_3(\mu_3\text{-}O)(\mu_2\text{-}OH)_3]_4(\mathbf{C2})_4$; (c) $[Cp_3Zr_3(\mu_3\text{-}O)(\mu_2\text{-}OH)_3]_4(\mathbf{D6})_4$; (d) $[Cp_3Zr_3(\mu_3\text{-}O)(\mu_2\text{-}OH)_3]_4(\mathbf{C3})_4$; (e) $[Cp_3Zr_3(\mu_3\text{-}O)(\mu_2\text{-}OH)_3]_4(\mathbf{C6})_4$; (f) $[Cp_3Zr_3(\mu_3\text{-}O)(\mu_2\text{-}OH)_3]_4(\mathbf{C7})_4$; (g) $[Cp_3Zr_3(\mu_3\text{-}O)(\mu_2\text{-}OH)_3]_4(\mathbf{C8})_4$.

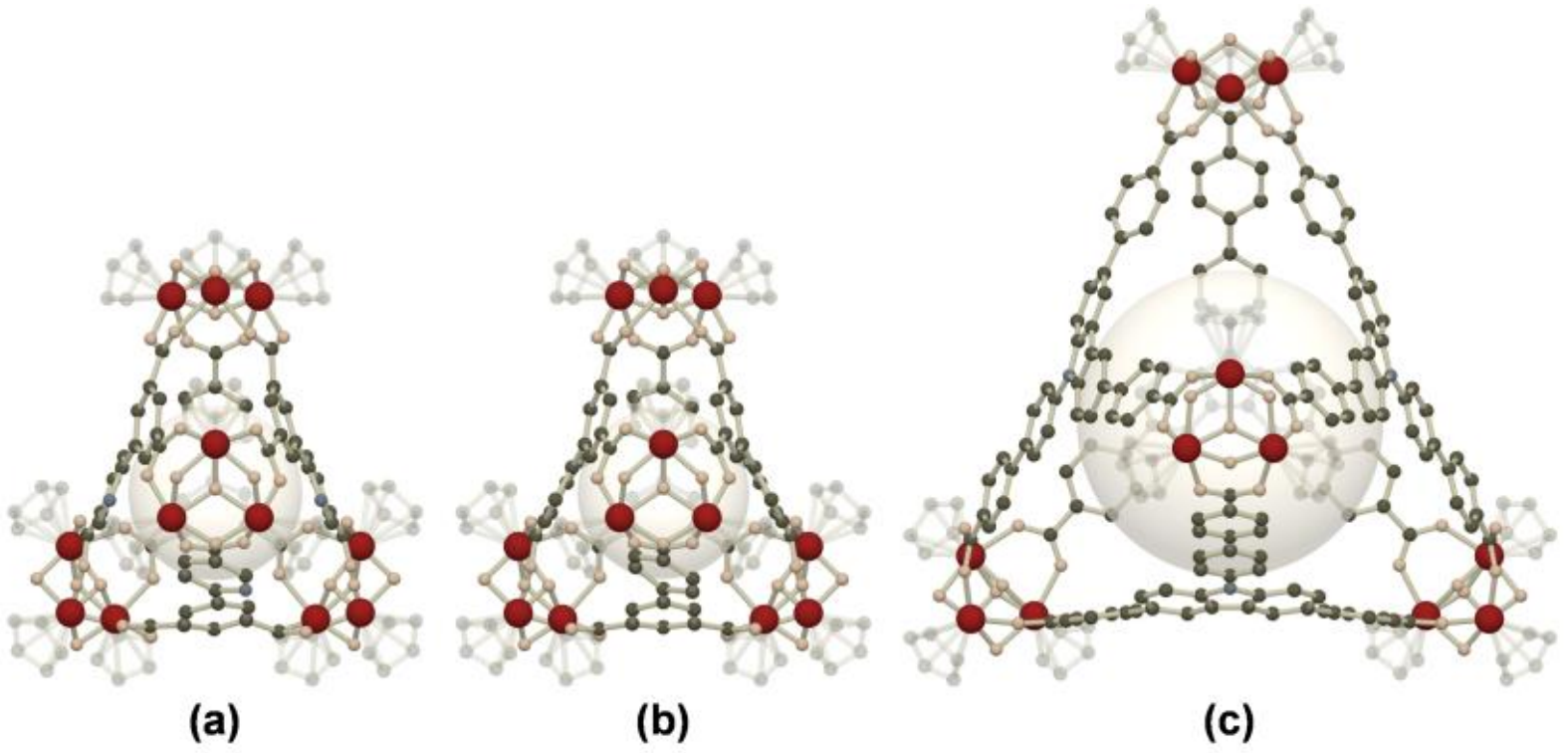

**Figure 20.** Structural illustrations of face-directed $M_4L_4$ tetrahedra, $[Cp_3Zr_3(\mu_3\text{-}O)(\mu_2\text{-}OH)_3]_4(\mathbf{L})_4$, built from tritopic linkers without three-fold symmetry. (a) $[Cp_3Zr_3(\mu_3\text{-}O)(\mu_2\text{-}OH)_3]_4(\mathbf{C13})_4$; (b) $[Cp_3Zr_3(\mu_3\text{-}O)(\mu_2\text{-}OH)_3]_4(\mathbf{C12})_4$; (c) $[Cp_3Zr_3(\mu_3\text{-}O)(\mu_2\text{-}OH)_3]_4(\mathbf{C14})_4$.

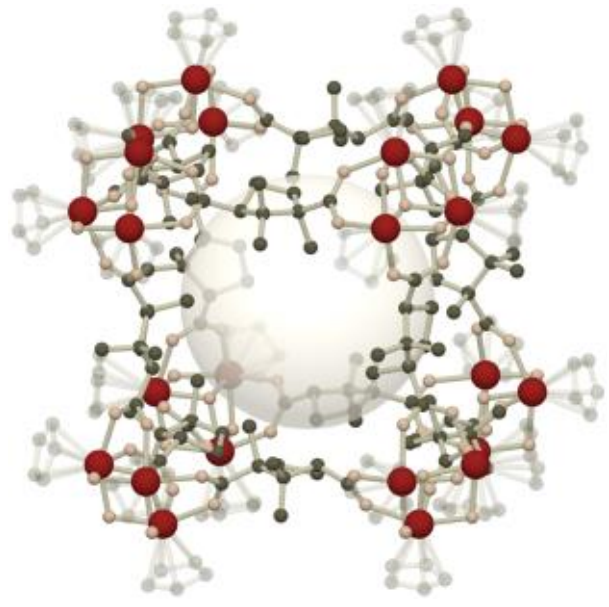

**Figure 21.** Structural illustration of an $M_8L_{12}$ cube. $[Cp_3Zr_3(\mu_3\text{-}O)(\mu_2\text{-}OH)_3]_8(\mathbf{B12})_{12}$.

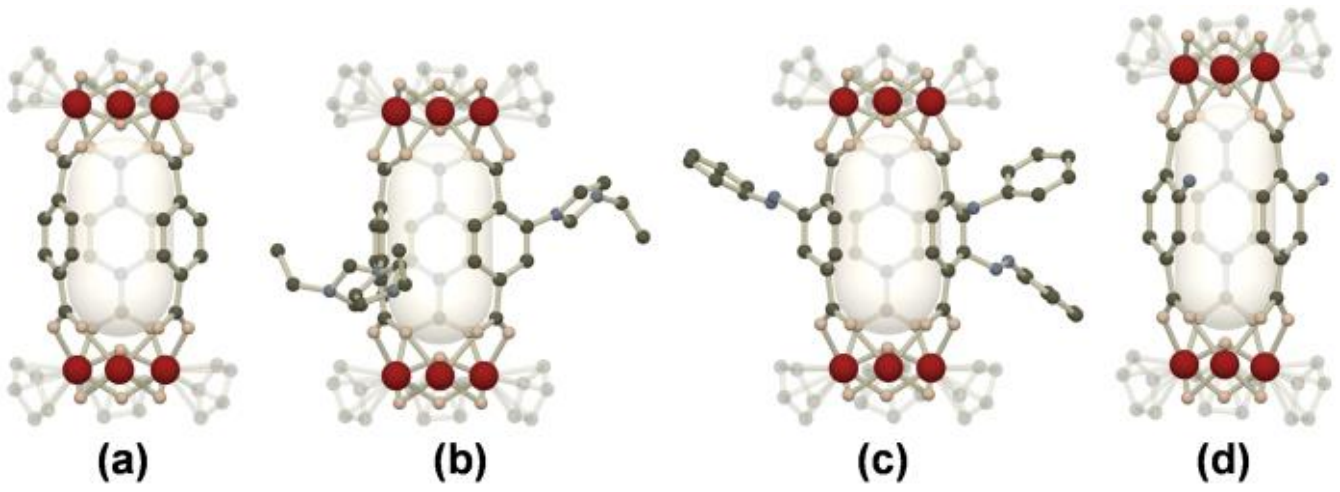

**Figure 22.** Structural illustrations of $M_2L_3$ cocoons, $[Cp_3Zr_3(\mu_3\text{-}O)(\mu_2\text{-}OH)_3]_2(\mathbf{L})_3$, built from BDC-based linkers. (a) $[Cp_3Zr_3(\mu_3\text{-}O)(\mu_2\text{-}OH)_3]_2(\mathbf{A1})_3$; (b) $[Cp_3Zr_3(\mu_3\text{-}O)(\mu_2\text{-}OH)_3]_2(\mathbf{A23})_3$; (c) $[Cp_3Zr_3(\mu_3\text{-}O)(\mu_2\text{-}OH)_3]_2(\mathbf{A25})_3$; (d) $[Cp_3Zr_3(\mu_3\text{-}O)(\mu_2\text{-}OH)_3]_2(\mathbf{A12})_3$.

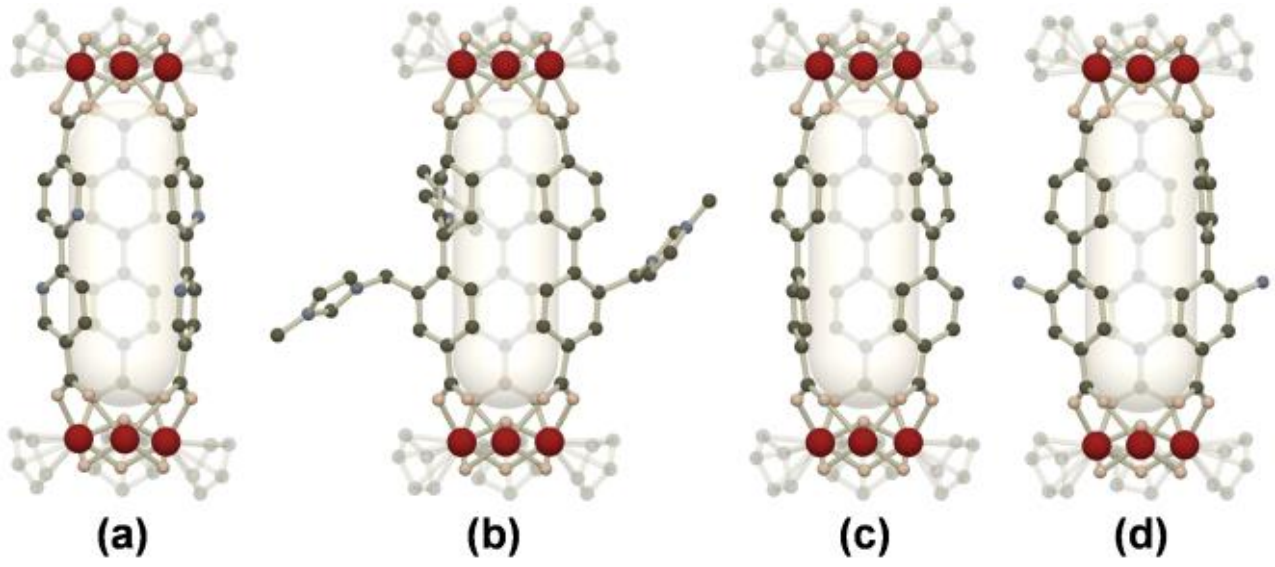

**Figure 23.** Structural illustrations of $M_2L_3$ cocoons, $[Cp_3Zr_3(\mu_3\text{-}O)(\mu_2\text{-}OH)_3]_2(\mathbf{L})_3$, built from BPDC-based linkers. (a) $[Cp_3Zr_3(\mu_3\text{-}O)(\mu_2\text{-}OH)_3]_2(\mathbf{A58})_3$; (b) $[Cp_3Zr_3(\mu_3\text{-}O)(\mu_2\text{-}OH)_3]_2(\mathbf{A55})_3$; (c) $[Cp_3Zr_3(\mu_3\text{-}O)(\mu_2\text{-}OH)_3]_2(\mathbf{A52})_3$; (d) $[Cp_3Zr_3(\mu_3\text{-}O)(\mu_2\text{-}OH)_3]_2(\mathbf{A54})_3$.

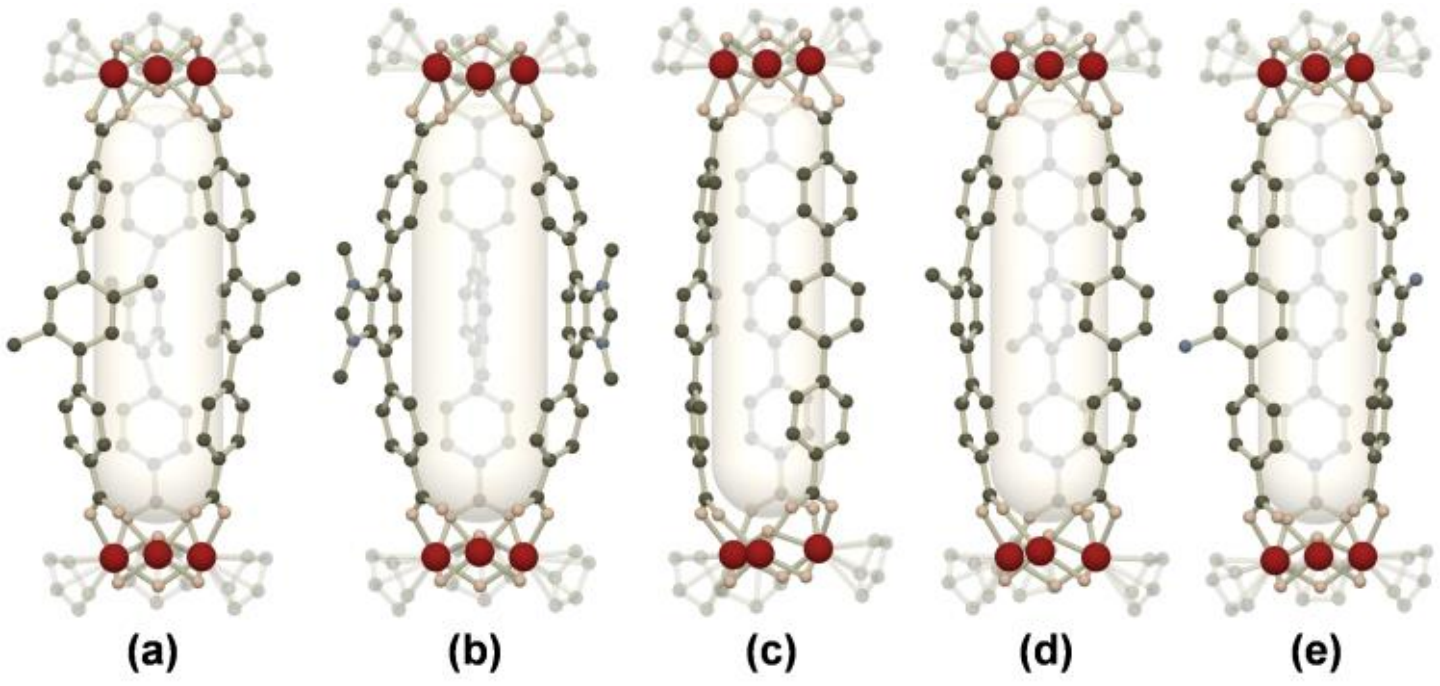

**Figure 24.** Structural illustrations of $M_2L_3$ cocoons, $[Cp_3Zr_3(\mu_3\text{-}O)(\mu_2\text{-}OH)_3]_2(\mathbf{L})_3$, built from TPDC-based linkers. (a) $[Cp_3Zr_3(\mu_3\text{-}O)(\mu_2\text{-}OH)_3]_2(\mathbf{A74})_3$; (b) $[Cp_3Zr_3(\mu_3\text{-}O)(\mu_2\text{-}OH)_3]_2(\mathbf{A78})_3$; (c) $[Cp_3Zr_3(\mu_3\text{-}O)(\mu_2\text{-}OH)_3]_2(\mathbf{A66})_3$; (d) $[Cp_3Zr_3(\mu_3\text{-}O)(\mu_2\text{-}OH)_3]_2(\mathbf{A67})_3$; (e) $[Cp_3Zr_3(\mu_3\text{-}O)(\mu_2\text{-}OH)_3]_2(\mathbf{A68})_3$.

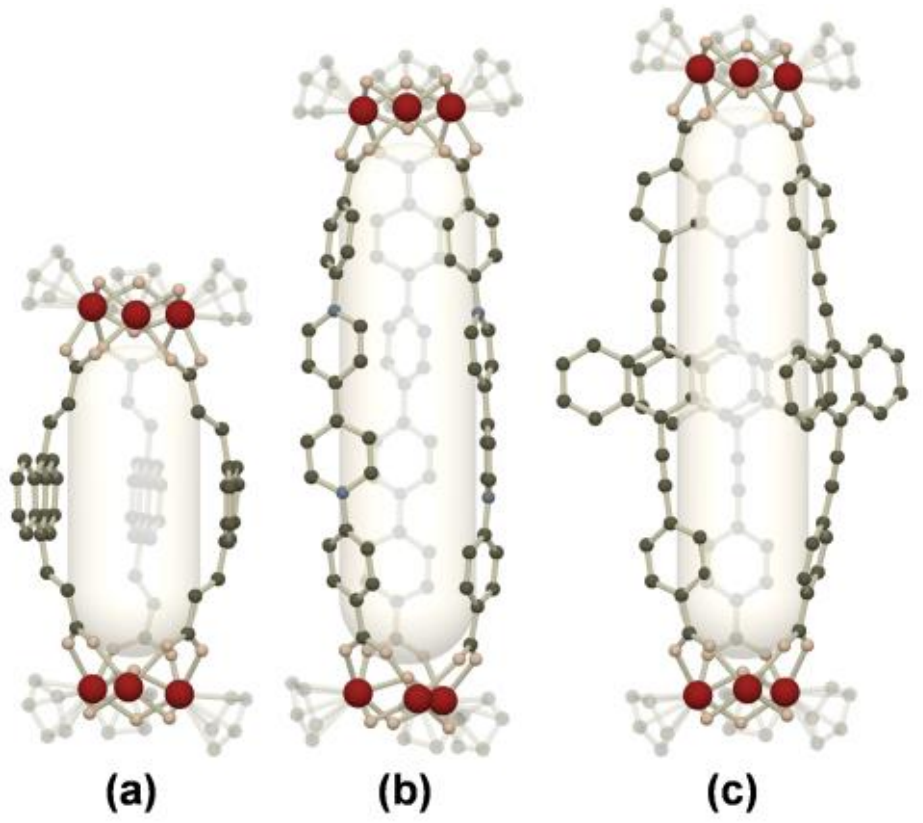

**Figure 25.** Structural illustrations of $M_2L_3$ cocoons, $[Cp_3Zr_3(\mu_3\text{-}O)(\mu_2\text{-}OH)_3]_2(\mathbf{L})_3$, built from miscellaneous and QPDC-based linear linkers. (a) $[Cp_3Zr_3(\mu_3\text{-}O)(\mu_2\text{-}OH)_3]_2(\mathbf{A92})_3$; (b) $[Cp_3Zr_3(\mu_3\text{-}O)(\mu_2\text{-}OH)_3]_2(\mathbf{A86})_3$; (c) $[Cp_3Zr_3(\mu_3\text{-}O)(\mu_2\text{-}OH)_3]_2(\mathbf{A93})_3$.

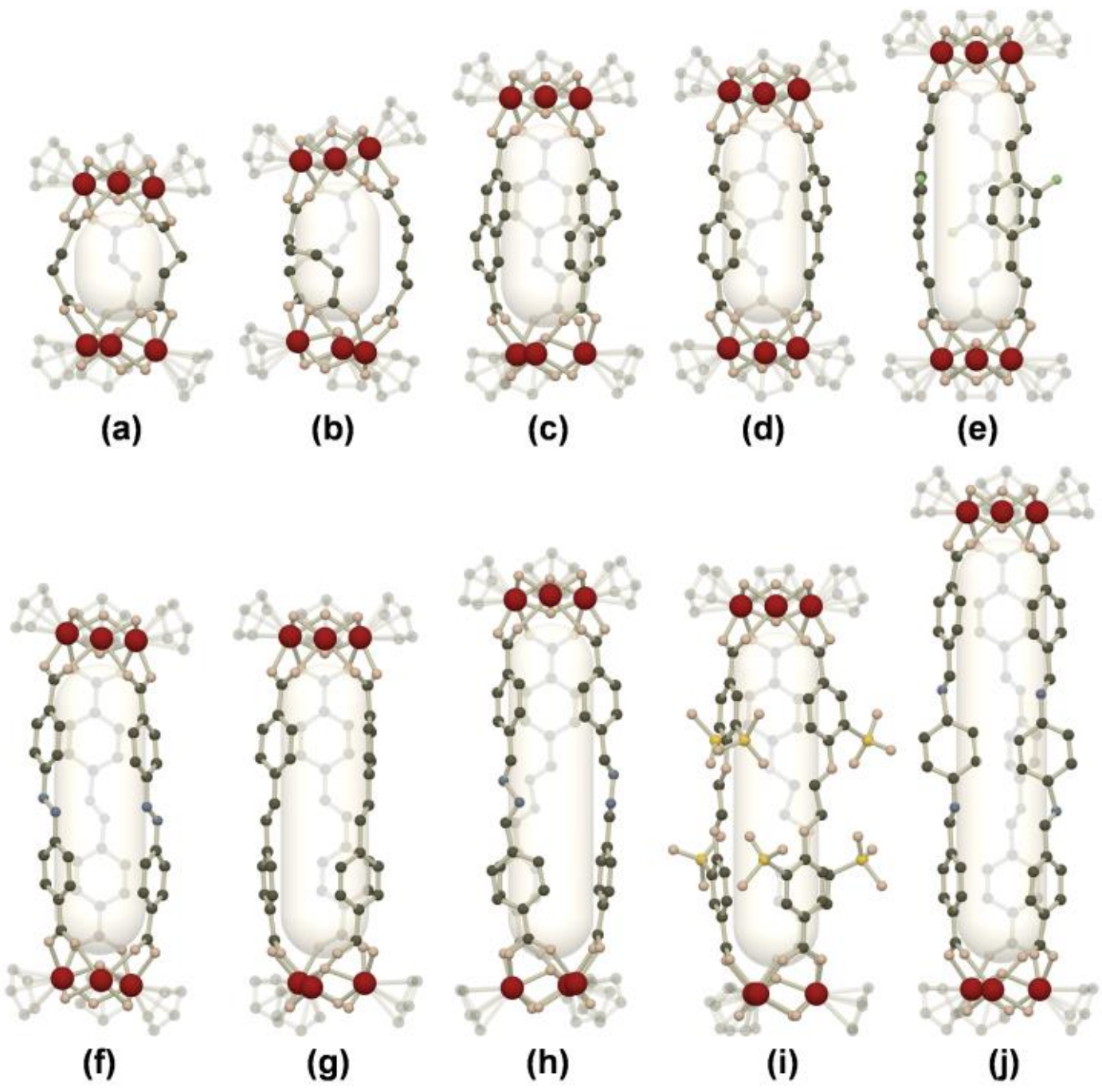

**Figure 26.** Structural illustrations of $M_2L_3$ cocoons, $[Cp_3Zr_3(\mu_3\text{-}O)(\mu_2\text{-}OH)_3]_2(\mathbf{L})_3$, built from zigzag-shaped linkers. (a) $[Cp_3Zr_3(\mu_3\text{-}O)(\mu_2\text{-}OH)_3]_2(\mathbf{A95})_3$; (b) $[Cp_3Zr_3(\mu_3\text{-}O)(\mu_2\text{-}OH)_3]_2(\mathbf{A96})_3$; (c) $[Cp_3Zr_3(\mu_3\text{-}O)(\mu_2\text{-}OH)_3]_2(\mathbf{A94})_3$; (d) $[Cp_3Zr_3(\mu_3\text{-}O)(\mu_2\text{-}OH)_3]_2(\mathbf{A97})_3$; (e) $[Cp_3Zr_3(\mu_3\text{-}O)(\mu_2\text{-}OH)_3]_2(\mathbf{A98})_3$; (f) $[Cp_3Zr_3(\mu_3\text{-}O)(\mu_2\text{-}OH)_3]_2(\mathbf{A101})_3$; (g) $[Cp_3Zr_3(\mu_3\text{-}O)(\mu_2\text{-}OH)_3]_2(\mathbf{A100})_3$; (h) $[Cp_3Zr_3(\mu_3\text{-}O)(\mu_2\text{-}OH)_3]_2(\mathbf{A102})_3$; (i) $[Cp_3Zr_3(\mu_3\text{-}O)(\mu_2\text{-}OH)_3]_2(\mathbf{A103})_3$; (j) $[Cp_3Zr_3(\mu_3\text{-}O)(\mu_2\text{-}OH)_3]_2(\mathbf{A104})_3$.

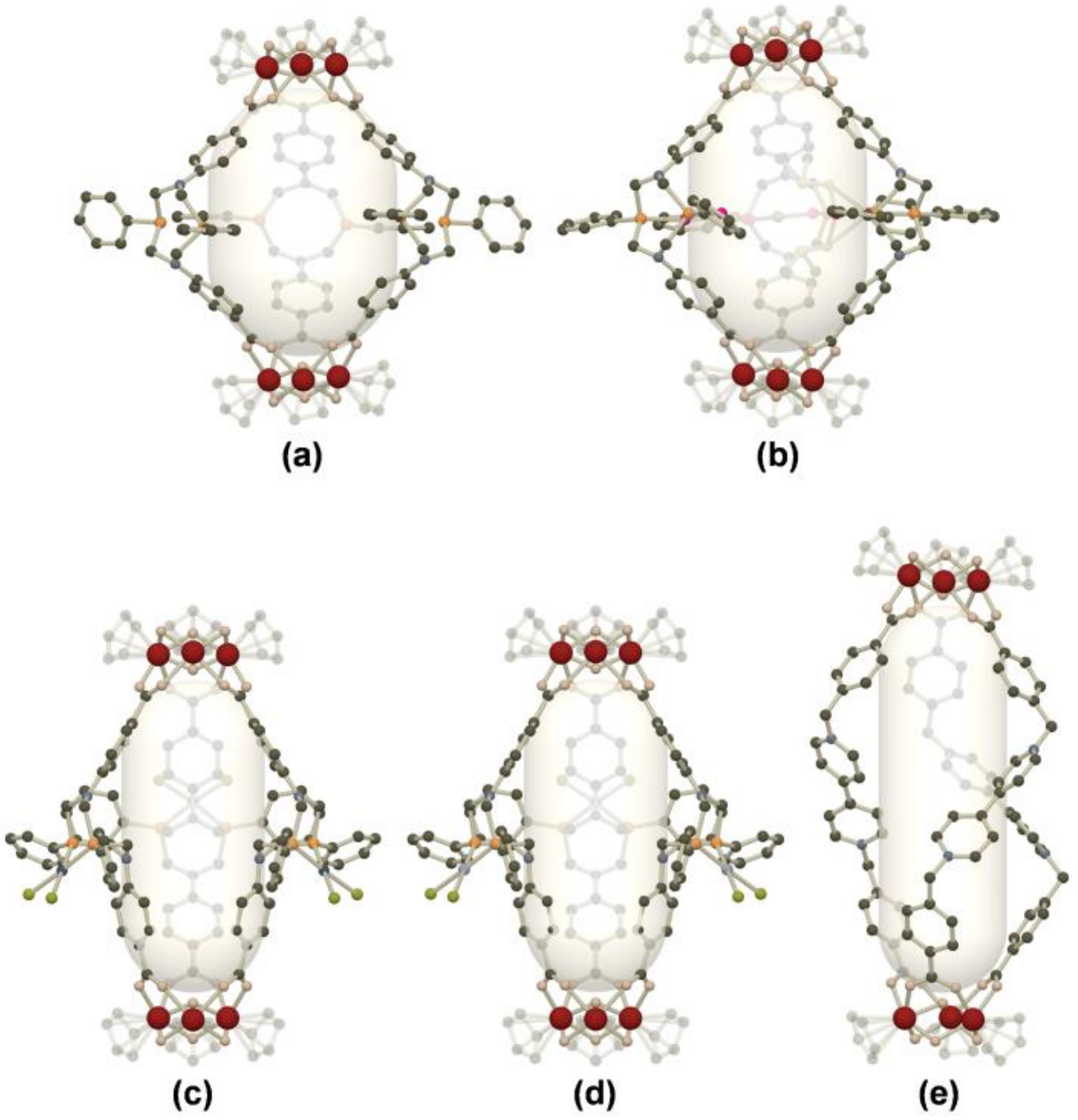

**Figure 27.** Structural illustrations of $M_2L_3$ cocoons, $[Cp_3Zr_3(\mu_3\text{-O})(\mu_2\text{-OH})_3]_2$(**L**)$_3$, built from zigzag-shaped linkers. (a) $[Cp_3Zr_3(\mu_3\text{-O})(\mu_2\text{-OH})_3]_2$(**A109**)$_3$; (b) $[Cp_3Zr_3(\mu_3\text{-O})(\mu_2\text{-OH})_3]_2$(**A112**)$_2$(**A113**); (c) $[Cp_3Zr_3(\mu_3\text{-O})(\mu_2\text{-OH})_3]_2$(**A110**)$_3$; (d) $[Cp_3Zr_3(\mu_3\text{-O})(\mu_2\text{-OH})_3]_2$(**A111**)$_3$; (e) $[Cp_3Zr_3(\mu_3\text{-O})(\mu_2\text{-OH})_3]_2$(**A107**)$_3$.

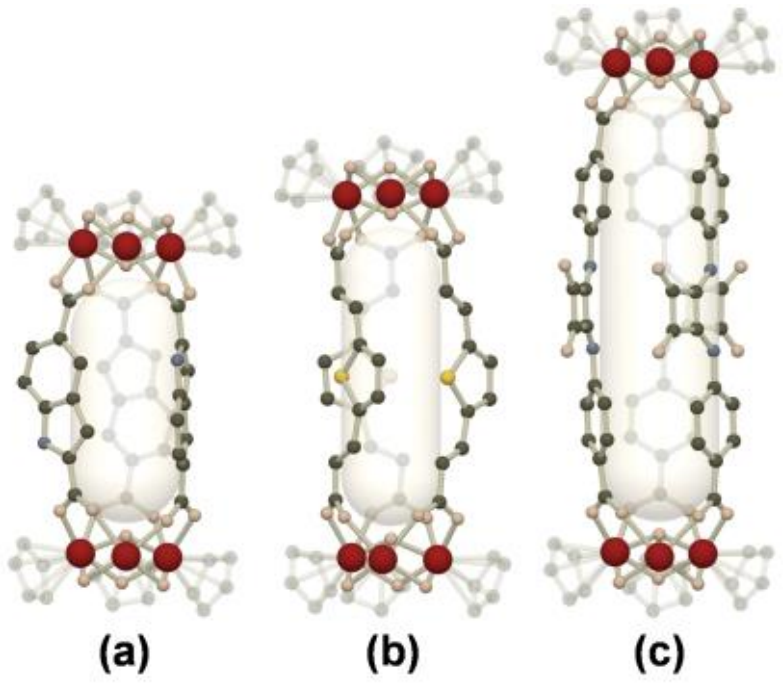

**Figure 28.** Structural illustrations of $M_2L_3$ cocoons, $[Cp_3Zr_3(\mu_3\text{-O})(\mu_2\text{-OH})_3]_2$(**L**)$_3$, built from bent linkers. (a) $[Cp_3Zr_3(\mu_3\text{-O})(\mu_2\text{-OH})_3]_2$(**B20**)$_3$; (b) $[Cp_3Zr_3(\mu_3\text{-O})(\mu_2\text{-OH})_3]_2$(**B13**)$_3$; (c) $[Cp_3Zr_3(\mu_3\text{-O})(\mu_2\text{-OH})_3]_2$(**B21**)$_3$.

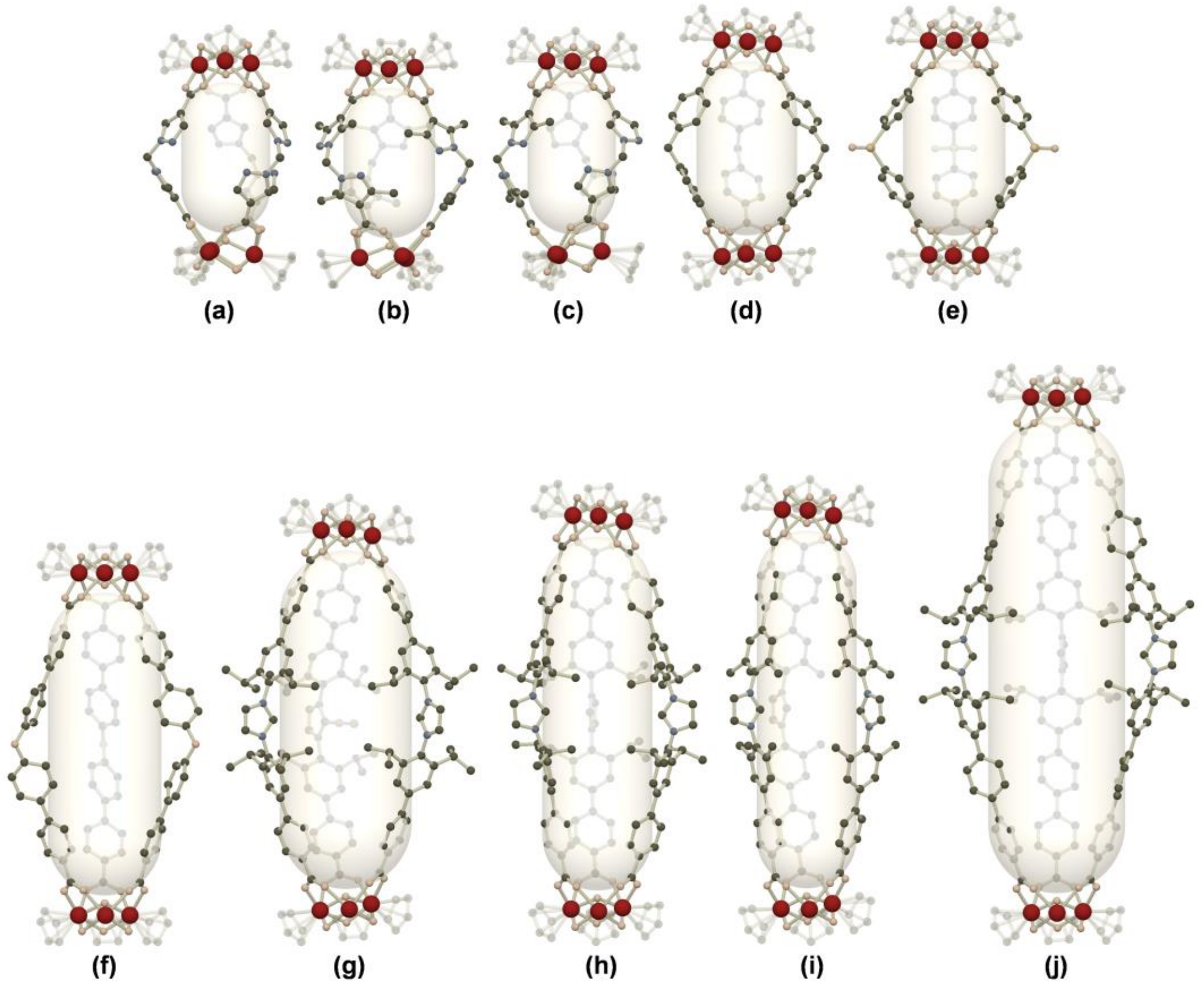

**Figure 29.** Structural illustrations of $M_2L_3$ cocoons, $[Cp_3Zr_3(\mu_3\text{-}O)(\mu_2\text{-}OH)_3]_2(\mathbf{L})_3$, built from bent linkers. (a) $[Cp_3Zr_3(\mu_3\text{-}O)(\mu_2\text{-}OH)_3]_2(\mathbf{B4})_3$; (b) $[Cp_3Zr_3(\mu_3\text{-}O)(\mu_2\text{-}OH)_3]_2(\mathbf{B5})_3$; (c) $[Cp_3Zr_3(\mu_3\text{-}O)(\mu_2\text{-}OH)_3]_2(\mathbf{B4})_2(\mathbf{B5})$; (d) $[Cp_3Zr_3(\mu_3\text{-}O)(\mu_2\text{-}OH)_3]_2(\mathbf{B7})_3$; (e) $[Cp_3Zr_3(\mu_3\text{-}O)(\mu_2\text{-}OH)_3]_2(\mathbf{B8})_3$; (f) $[Cp_3Zr_3(\mu_3\text{-}O)(\mu_2\text{-}OH)_3]_2(\mathbf{B9})_3$; (g) $[Cp_3Zr_3(\mu_3\text{-}O)(\mu_2\text{-}OH)_3]_2(\mathbf{B16})_2(\mathbf{B19})$; (h) $[Cp_3Zr_3(\mu_3\text{-}O)(\mu_2\text{-}OH)_3]_2(\mathbf{B16})_3$; (i) $[Cp_3Zr_3(\mu_3\text{-}O)(\mu_2\text{-}OH)_3]_2(\mathbf{B15})_3$; (j) $[Cp_3Zr_3(\mu_3\text{-}O)(\mu_2\text{-}OH)_3]_2(\mathbf{B17})_3$.

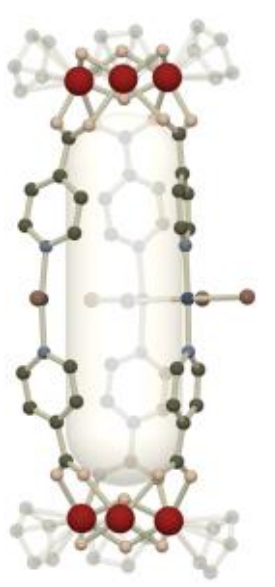

**Figure 30.** Structural illustration of an $M_2L_3$ cocoon, $[Cp_3Zr_3(\mu_3\text{-}O)(\mu_2\text{-}OH)_3]_2(\mathbf{L})_3$, built from a Pd-bridged ditopic linker. $[Cp_3Zr_3(\mu_3\text{-}O)(\mu_2\text{-}OH)_3]_2(\mathbf{D3})_3$.

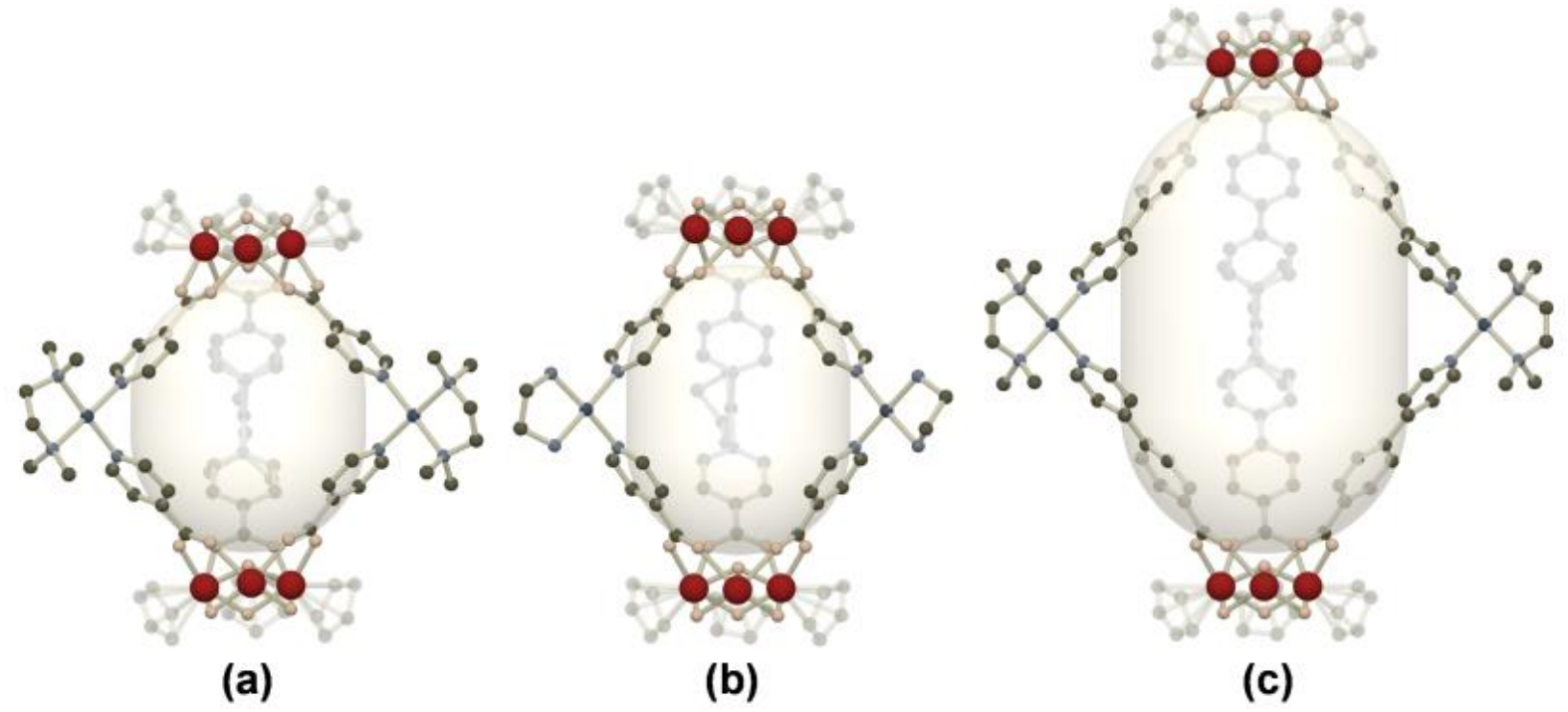

**Figure 31.** Structural illustrations of $M_2L_3$ cocoons, $[Cp_3Zr_3(\mu_3\text{-}O)(\mu_2\text{-}OH)_3]_2(\mathbf{L})_3$, built from Pd-bridged ditopic linkers. (a) $[Cp_3Zr_3(\mu_3\text{-}O)(\mu_2\text{-}OH)_3]_2(\mathbf{D2})_3$; (b) $[Cp_3Zr_3(\mu_3\text{-}O)(\mu_2\text{-}OH)_3]_2(\mathbf{D1})_3$; (c) $[Cp_3Zr_3(\mu_3\text{-}O)(\mu_2\text{-}OH)_3]_2(\mathbf{D5})_3$.

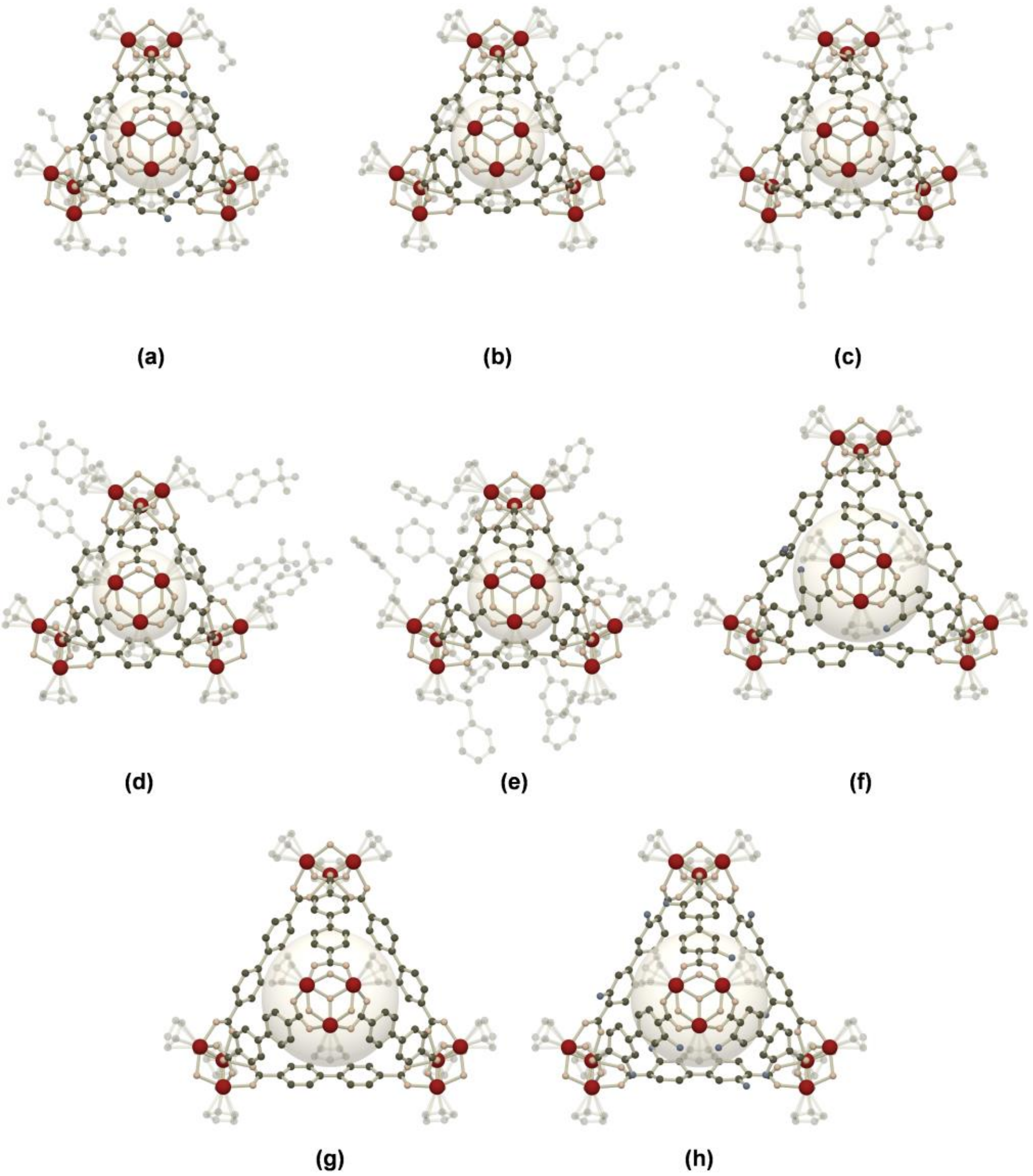

**Figure 32.** Structural illustrations of edge-directed $M_4L_6$ tetrahedra, $[(RCp)_3Zr_3(\mu_3\text{-}O)(\mu_2\text{-}OH)_3]_4(\mathbf{L})_6$, capped by substituted cyclopentadienyl (RCp) ligands. (a) $[(n\text{BuCp})_3Zr_3(\mu_3\text{-}O)(\mu_2\text{-}OH)_3]_4(\mathbf{A12})_6$; (b) $[(4\text{-vinylbenzyl-Cp})_3Zr_3(\mu_3\text{-}O)(\mu_2\text{-}OH)_3]_4(\mathbf{A1})_6$; (c) $[(n\text{BuCp})_3Zr_3(\mu_3\text{-}O)(\mu_2\text{-}OH)_3]_4(\mathbf{A1})_6$; (d) $[(4\text{-}CF_3\text{benzyl-Cp})_3Zr_3(\mu_3\text{-}O)(\mu_2\text{-}OH)_3]_4(\mathbf{A1})_6$; (e) $[(\text{benzyl-Cp})_3Zr_3(\mu_3\text{-}$

O)($\mu_2$-OH)$_3$]$_4$(**A1**)$_6$; (f) [($n$BuCp)$_3$Zr$_3$($\mu_3$-O)($\mu_2$-OH)$_3$]$_4$(**A54**)$_6$; (g) [($n$BuCp)$_3$Zr$_3$($\mu_3$-O)($\mu_2$-OH)$_3$]$_4$(**A52**)$_6$; (h) [($n$BuCp)$_3$Zr$_3$($\mu_3$-O)($\mu_2$-OH)$_3$]$_4$(**A56**)$_6$.

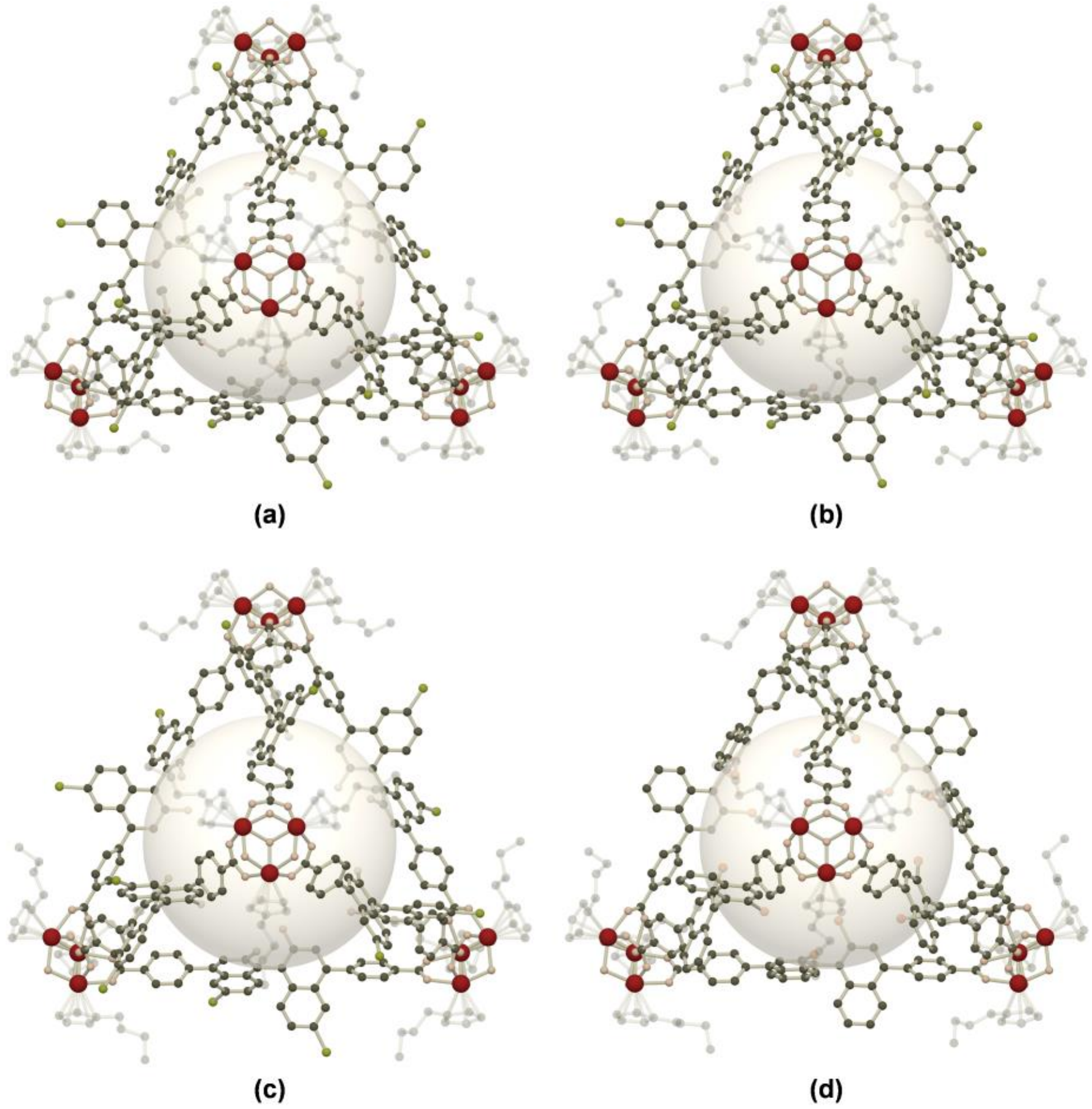

**Figure 33.** Structural illustrations of edge-directed $M_4L_6$ tetrahedra, [(RCp)$_3$Zr$_3$($\mu_3$-O)($\mu_2$-OH)$_3$]$_4$(**L**)$_6$, capped by substituted cyclopentadienyl (RCp) ligands. (a) [($n$BuCp)$_3$Zr$_3$($\mu_3$-O)($\mu_2$-OH)$_3$]$_4$(**A82**)$_6$; (b) [($n$BuCp)$_3$Zr$_3$($\mu_3$-O)($\mu_2$-OH)$_3$]$_4$(**A83**)$_6$; (c) [($n$BuCp)$_3$Zr$_3$($\mu_3$-O)($\mu_2$-OH)$_3$]$_4$(**A81**)$_6$; (d) [($n$BuCp)$_3$Zr$_3$($\mu_3$-O)($\mu_2$-OH)$_3$]$_4$(**A80**)$_6$.

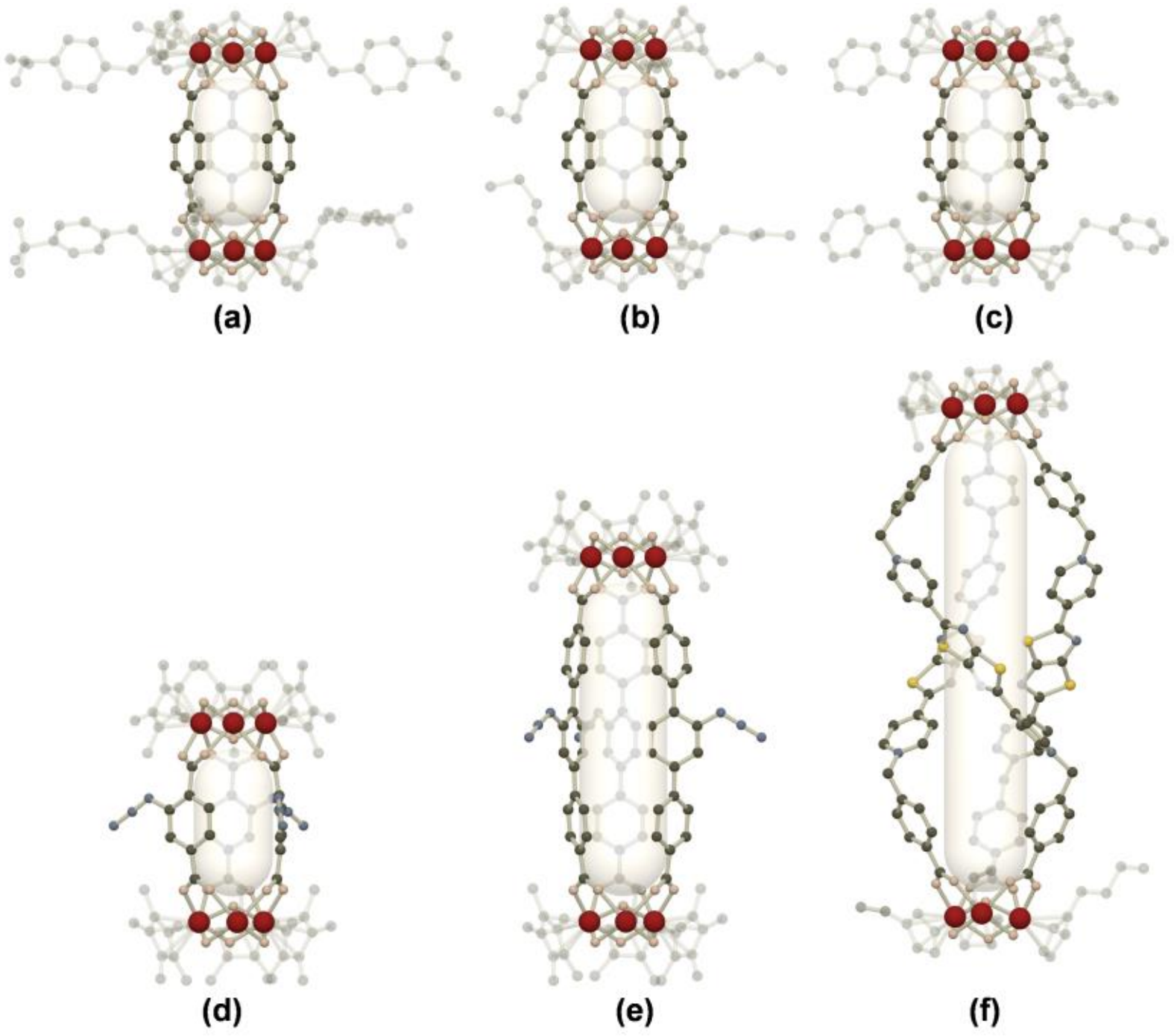

**Figure 34.** Structural illustrations of $M_2L_3$ cocoons, $[(RCp)_3Zr_3(\mu_3\text{-}O)(\mu_2\text{-}OH)_3]_2(\mathbf{L})_3$, capped by substituted cyclopentadienyl (RCp) ligands. (a) $[(4\text{-}CF_3benzyl\text{-}Cp)_3Zr_3(\mu_3\text{-}O)(\mu_2\text{-}OH)_3]_2(\mathbf{A1})_3$; (b) $[(n\text{BuCp})_3Zr_3(\mu_3\text{-}O)(\mu_2\text{-}OH)_3]_2(\mathbf{A1})_3$; (c) $[(benzyl\text{-}Cp)_3Zr_3(\mu_3\text{-}O)(\mu_2\text{-}OH)_3]_2(\mathbf{A1})_3$; (d) $[Cp^*{}_3Zr_3(\mu_3\text{-}O)(\mu_2\text{-}OH)_3]_2(\mathbf{A14})_3$; (e) $[Cp^*{}_3Zr_3(\mu_3\text{-}O)(\mu_2\text{-}OH)_3]_2(\mathbf{A69})_3$; (f) $[(n\text{BuCp})_3Zr_3(\mu_3\text{-}O)(\mu_2\text{-}OH)_3]_2(\mathbf{A108})_3$.

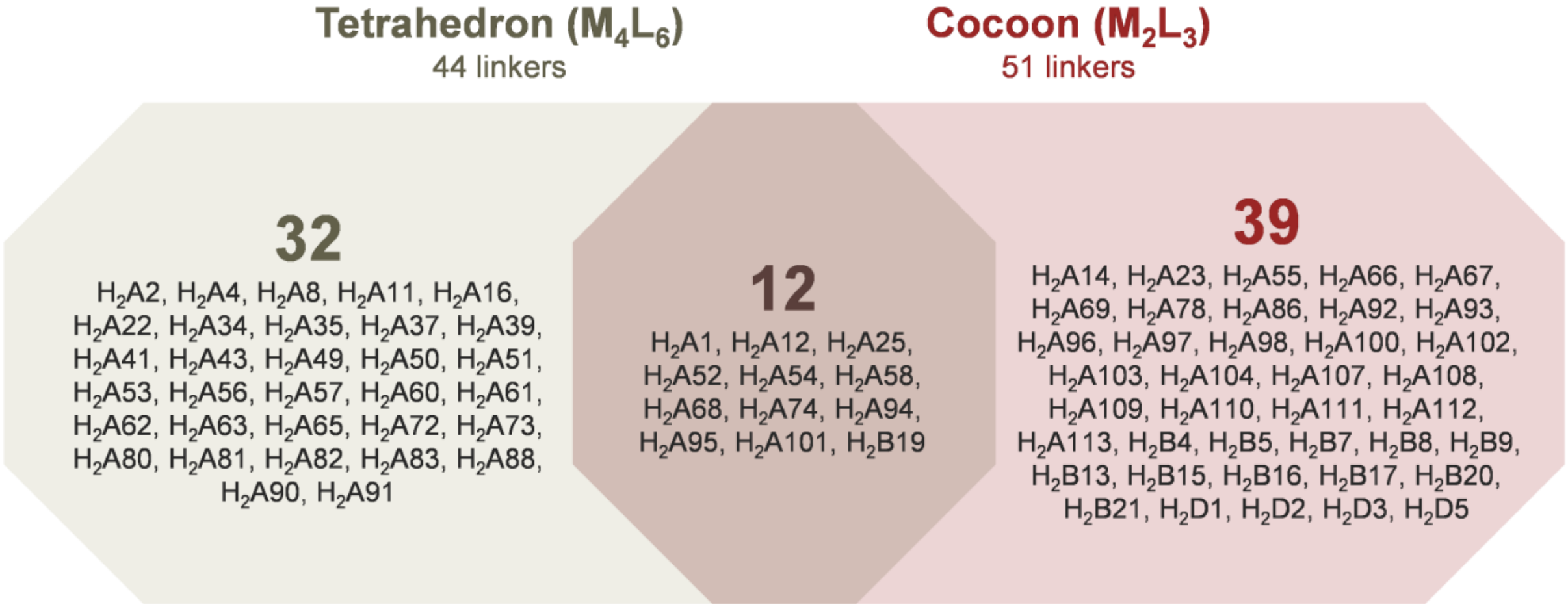

**Figure 35.** Ditopic (2-c) linkers with a reported crystallographic information file (CIF) in Zr-MOPs assembled from Cp-capped $Zr_3$ clusters, grouped by the MOP types they form. Linkers in the overlapping region have CIFs for both types. Linkers from mixed-linker MOPs are included under the corresponding MOP type.

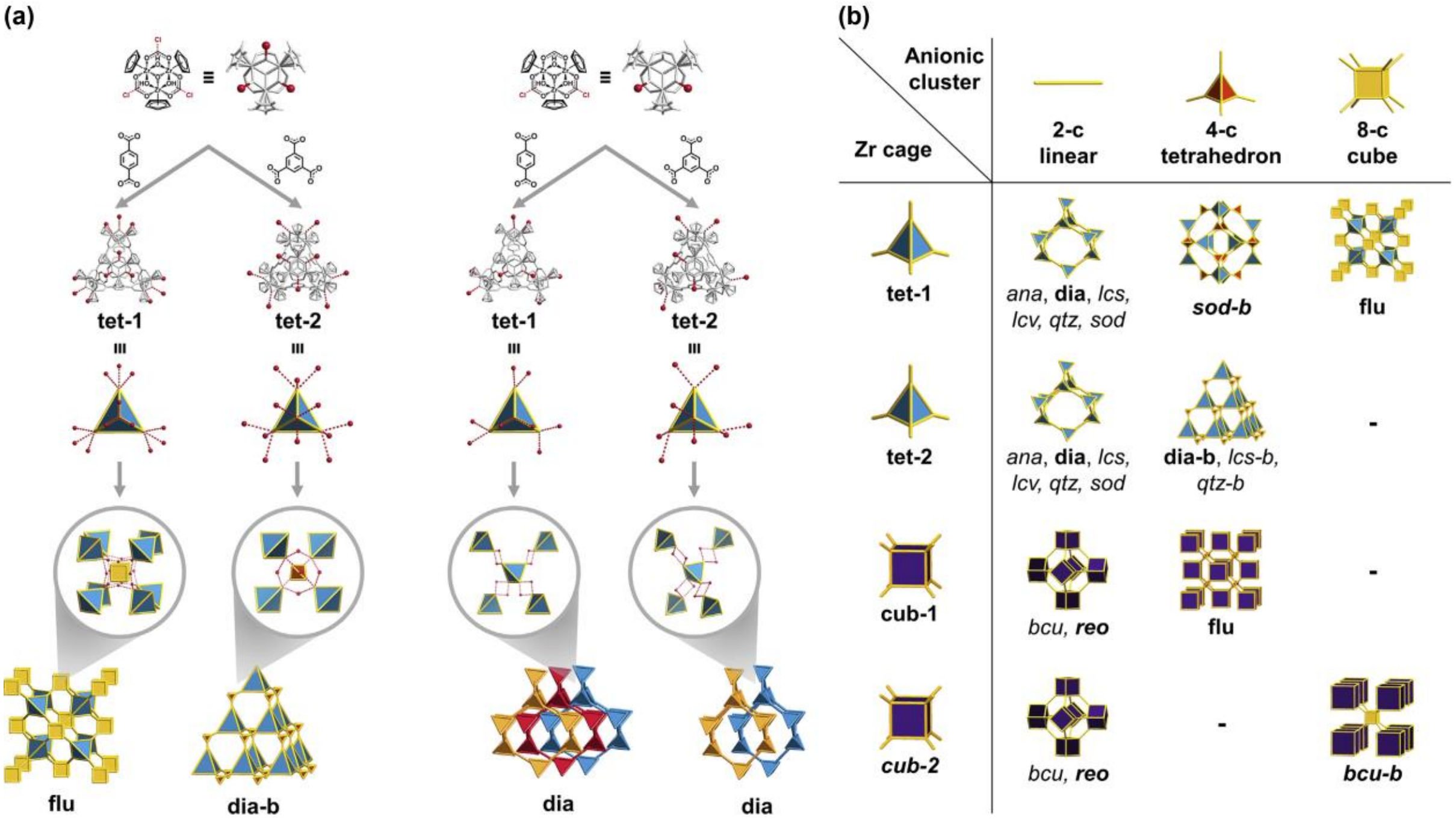

**Figure 36.** Packing of Zr-based cages through hydroxo–anion contacts. (a) Ditopic and tritopic linkers orient the $\mu_2$-OH groups of the $Zr_3$ nodes toward the edges or the faces of the tetrahedron. When all three hydroxo groups bind chloride, the cages give **flu** or **dia-b** nets; when two bind, both cage types give interpenetrated **dia** nets. (b) Reticular table of reported and hypothetical nets for each combination of Zr cage and anionic cluster. Reported nets are shown in bold, and hypothetical nets and the unknown cage (***cub-2***) in italics; **tet-1** and **tet-2** correspond to edge- and face-directed tetrahedra, respectively.[25] Adapted from *Chem*, 8, J. Kim and W. Choe, "Topology-guided roadmap for reticular chemistry of metal-organic polyhedra," 617–631, Copyright (2022), with permission from Elsevier.

## Table of Contents

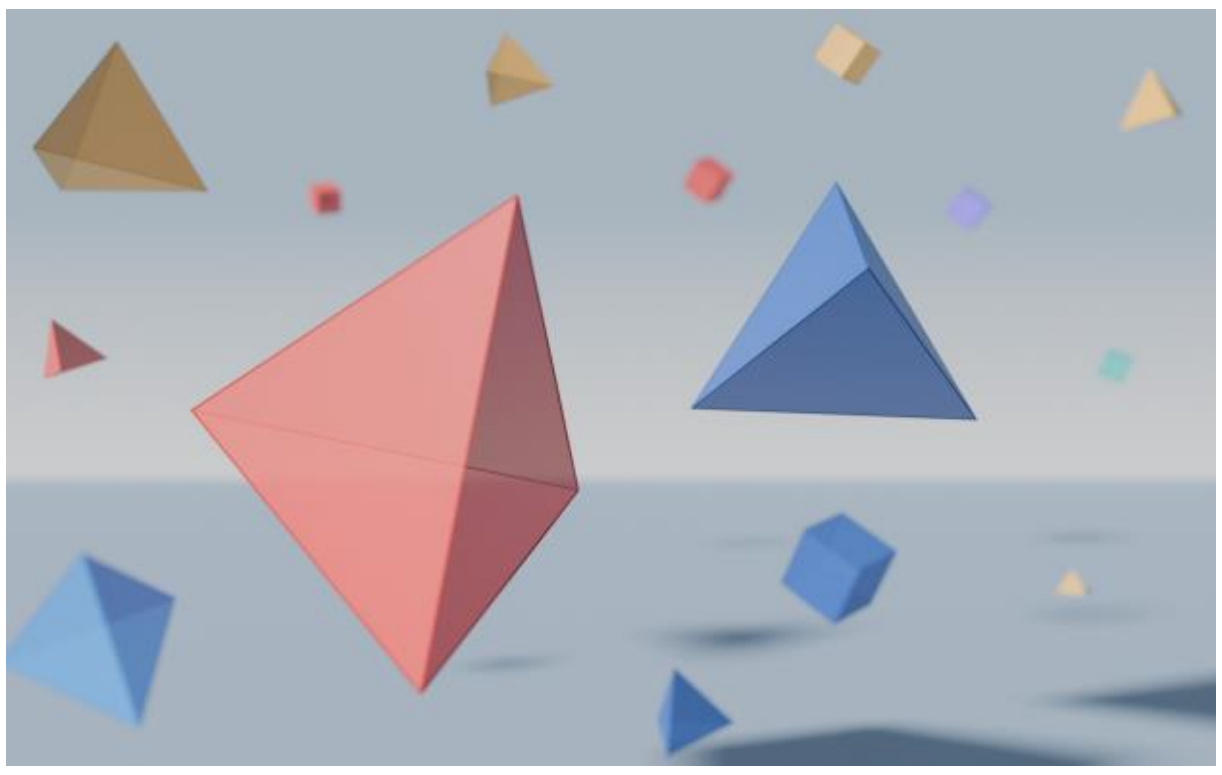

Zirconium metal–organic polyhedra are discrete molecular cages that become building units of larger assemblies in the solid state. This review surveys their structures and examines how building blocks, assembly conditions, and intermolecular interactions shape cage formation and crystal packing, providing a guide for future design.